# Elasticity-Driven Periodic Polarization Patterns in Confined Chiral Ferroelectric Nematic Fluid

Anej Sterle[1,2], Peter Medle-Rupnik[1,2], Luka Cmok[1,2,3], Aitor Erkoreka,[4] Marta Lavrič[1], Natan Osterman[1,2], Calum J. Gibb[5], J. Hobbs[6], Josu Martinez-Perdiguero,[4] Richard J. Mandle[5,6], Alenka Mertelj[1], Nerea Sebastián[1*]

[1] Jožef Stefan Institute, Ljubljana, Slovenia
[2] Faculty of Mathematics and Physics, University of Ljubljana, Ljubljana, Slovenia
[3] CENN Nanocenter, Ljubljana, Slovenia
[4] Department of Physics, Faculty of Science and Technology, University of the Basque Country UPV/EHU, Bilbao, Spain
[5] School of Chemistry, University of Leeds, Leeds, UK
[6] School of Physics and Astronomy, University of Leeds, Leeds, UK

*Corresponding author: nerea.sebastian@ijs.si

Keywords: Ferroelectric nematic fluids, polarization modulations, elastic instabilities, second harmonic generation, confinement effects

Ferroelectric nematic phases are a new class of polar fluids in which spontaneous polarization is directly coupled to the orientational order, providing unique opportunities for creating self-organized materials with spatially modulated electric polarization and nonlinear optical response. Here we report the spontaneous emergence of polarization-modulated textures in a chiral ferroelectric nematic material close to the transition to the chiral twist-bend ferroelectric nematic phase. By systematically varying cell thickness and surface anchoring conditions, we map the formation of these modulated states, revealing stripe, square, and hexagonal morphologies determined via confinement conditions. These structures are directly translated into periodic modulation of the nonlinear optical response, as evidenced by second harmonic generation imaging. Comparison with an elasticity based theoretical framework and numerical free energy minimization shows that the instability originates from the softening of the bend elastic constant in the chiral nematic phase as the system approaches the lower temperature heliconical polar phase. The resulting elastic frustration, combined with confinement, drives the formation of spatially periodic director distortions, highlighting ferroelectric nematic fluids as a promising platform for self-assembled nonlinear optical materials.

## 1. Introduction

Self-assembly in soft matter provides a powerful route to functional materials with complex, hierarchically organized structures. Different mechanisms operate over different interaction ranges, resulting in a broad range of self-assembly length scales, from nanoscales, like in DNA,[1] micelles and block copolymers,[2,3], to microscales, e.g. in liquid crystals.[4] The latter are particularly attractive due to their intrinsic birefringence and strong responsiveness to external stimuli, enabling the formation of hierarchical optical architectures with potential for advanced photonic technologies.[5–8]

Chiral nematic liquid crystals (N*) form a self-organized helical structure, in which the average molecular orientation, i.e., the director **n**, twists periodically, with the distance for a full 360° rotation defined as the pitch, *p*. These systems exhibit a variety of instabilities under different conditions, such as the well-known Helfrich-Hurault periodic pseudolayer undulations under applied electric or magnetic fields[9–12] as well as driven via geometric frustration,[13–15] finger-like instabilities via surface anchoring[16] or light driven 3D manipulation of the helical axis promoting 1D and 2D structures.[17] Heliconical instabilities under electric or magnetic fields in N* systems in which the bend elastic constant is lower than the twist ($K_3$<$K_2$), were predicted by de P. Gennes[18] and R.B. Meyer.[19] The early 2010s discovery of the twist-bend nematic phase in odd flexible dimers[20–22] featuring low values of $K_3$, enabled experimental realization.[23,24] In these systems, it was subsequently shown that geometric confinement frustration arising from the combination of thickness and surface alignment constraints acts as an elastic field, spontaneously forming undulation instabilities with different directions and dimensionalities.[25–28]

In recent years, research on liquid crystalline phases has undergone a profound transformation, with the discovery of the ferroelectric nematic counterparts: the ferroelectric nematic phase ($N_F$)[29–32] or the chiral ferroelectric nematic phase ($N_F$*).[33–35] In contrast to the apolar conventional nematic phase, these ferroelectric nematic phases are polar, exhibiting a spontaneous macroscopic polarization that is spatially uniform in $N_F$ and helically twisted in $N_F$*. Self-assembled periodic patterns in polar nematic fluids, which exhibit large nonlinear optical coefficients,[36,37] are of interest both from a fundamental perspective, providing an experimentally accessible platform for exploring polarization topology, and for intrinsic regular modulation of nonlinear susceptibilities in reconfigurable materials. Recently, a novel twist-bend ferroelectric nematic phase ($N_{TBF}$*) has been identified.[38] The same mesophase was identified for a system of hard-rod polar molecular motives,[39] tagged $^{HC}N_F$, and in systems with the presence of cybotactic smectic correlations, addressed as HEC mesophase.[40] Similarly, a layered phase with heliconical polarization, termed $SmC^P_H$, has also been reported.[41] All of them share a heliconical and polar molecular arrangement, making them both ferroelectric and chiral, with a periodicity in the range of visible light wavelengths.

Here, we design a chiral ferroelectric material showing a room temperature $N^*_{TBF}$ phase and demonstrate the spontaneous formation of elasticity-driven modulated structures, occurring in the vicinity of the transition from the $N_F$* to the $N_{TBF}$* phase. The appearance of these modulated polarization architectures is controlled and investigated as a function of confining thicknesses and surface anchoring geometry. By comparison with an elasticity-driven model and numerical free-energy minimization, we show that these instabilities arise from the combined effect of bend elastic constant softening in the nematic phase and confinement conditions as the system approaches the lower-temperature $N_{TBF}$* phase. The calculated elasticity-driven polarization self-assembled structures show low electrostatic cost, with electrostatic self-interaction not playing a dominant role, highlighting an avenue for effective geometric control of 3D polarization patterning.

## 2. Results and Discussion

### *2.1 Material and mesophase behavior*

We designed the binary mixture F7, composed of 70% DIO [29] and 30% Compound 1 (C1)[41] (Figure 1), in order to stabilize the $N_F$ phase over an extended temperature range, followed by a narrow room-temperature $N_{TBF}$ phase. The heat-capacity profile obtained by high-resolution AC calorimetry on cooling is shown in Figure 1.d. The nematic-antiferroelectric nematic (N-$N_S$ ,99.4 °C) and the antiferroelectric nematic to ferroelectric nematic ($N_S$-$N_F$, 88.8 °C) phase transitions exhibit single and sharp peaks, demonstrating an excellent miscibility of both components. The negative peaks of the phase shift for both transitions (Figure S2) indicate second-order phase transitions, in contrast to pure DIO, where both are weakly first order.[42] Phase assignment of these three mesophases is supported by SAXS, POM, SHG and spontaneous polarization measurements (Figures S2 and S3), revealing a ferroelectric nematic phase extending from 88.8 °C to almost room temperature.

Around 29 °C, a clear textural transition is observed by POM (Figure 1.e, Movie S1), accompanied by visible light diffraction along the rubbing direction (Figure 1.c), indicating the emergence of a periodic submicron scale structure, not resolved by POM. Using a 448 nm source, diffraction peaks emerge at nearly 40 deg at the transition and rapidly shift to higher angles, suggesting an initial periodicity of the order of ~700 nm, which decreases quickly on cooling. X-ray experiments confirm a nematic pattern (Figure 1.f), identifying the mesophase as $N_{TBF}$. On further cooling, a transition at 21 °C (Figure 1.e) shows a grainy uniform texture with SAXS revealing a narrowed shifted peak (Figure 1.f and Figure S3), indicating a transition to a tilted smectic phase. At lower temperatures, calorimetry shows a broad anomaly with long $c_p$ pretransitional wings. Although two separate transitions cannot be distinguished reliably (i.e., as two separate peaks) as in the case of POM, the possibility of two broad transitions occurring close to each other cannot be excluded from the $c_p$ profile either. The tilted smectic phase additionally exhibits selective reflection (Figure 1.c), spontaneous polarization, as well as SHG signal (Figure S4). Thus, the mesophase is assigned to be a tilted smectic polar heliconical phase, $SmC^P_H$. Further analysis of the mesophases is beyond the scope of this manuscript, and here we will focus on the $N_F$-$N_{TBF}$ phase transition in the F7 chiral counterpart.

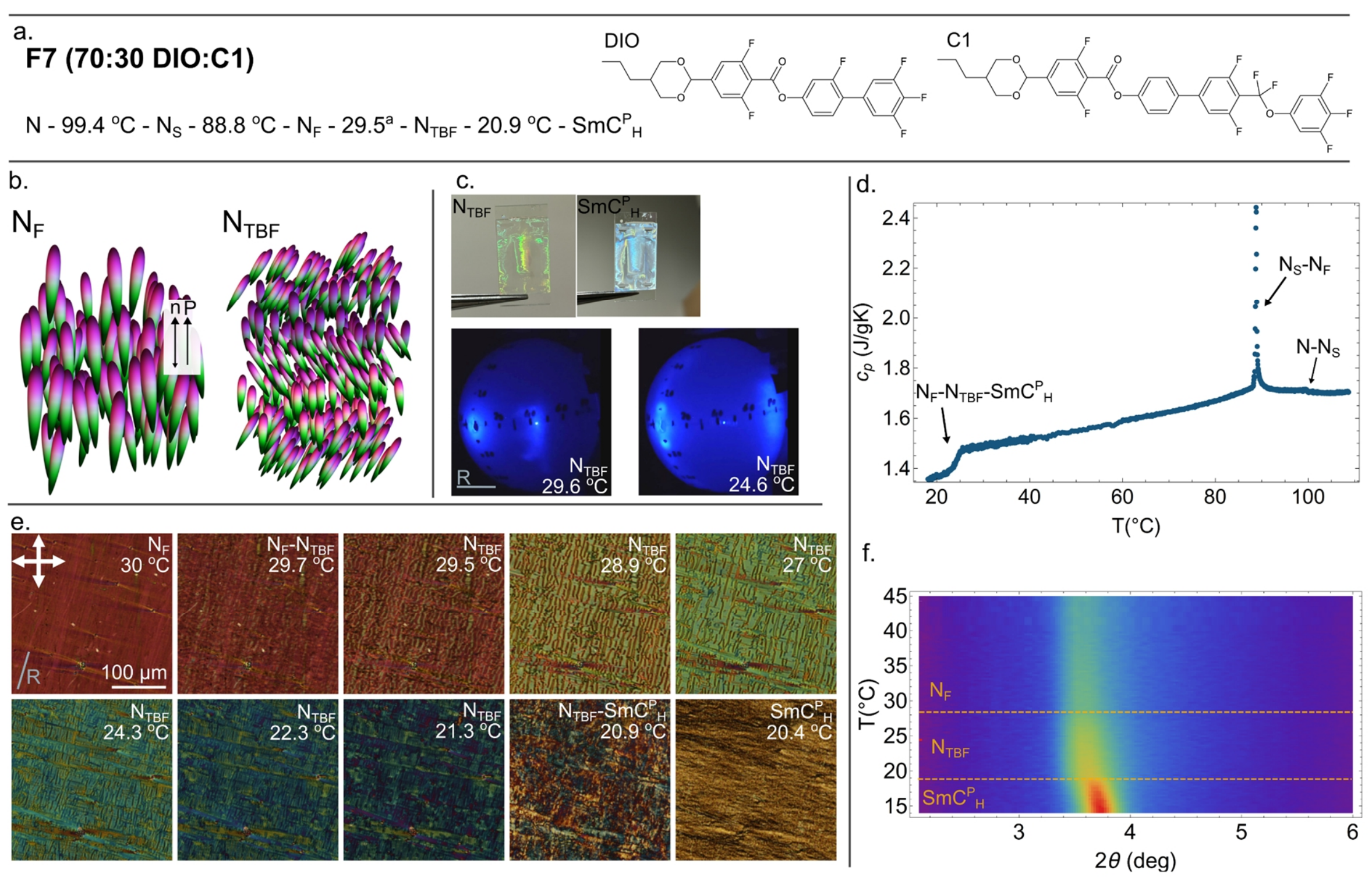

**Figure 1. Room temperature ferroelectric twist-bend nematic material.** a) F7 binary mixture phase behaviour on cooling and chemical structure of the mixture components DIO and C1. b) Schematic representation of the $N_F$ and $N_{TBF}$ phases. c) Visible reflected light perpendicularly to the rubbing direction and light diffraction experiment at the onset of the $N_{TBF}$ phase on 2 µm parallel rubbed cell, illuminated from behind and detected in a spherical screen. d) Temperature profile of the heat capacity, $c_p(T)$, upon cooling. e) $N_F$-$N_{TBF}$-$SmC^P_H$ phase sequence as seen by polarizing optical microscopy (POM) images on cooling in a 5 µm parallel rubbed cell. Double-headed arrows indicate the direction of crossed polarizers, and the light-blue line marked with R corresponds to the cell rubbing direction. f) Temperature evolution of the X-ray diffractograms obtained by integration of the 2D Broad-angle XRD for an unaligned sample in the vicinity of the $N_F$-$N_{TBF}$-$SmC^P_H$ phase transitions.

Building on the F7 mixture, we introduce chirality by adding 1wt% of the chiral dopant S811 (Merck) to F7. F7*exhibits a chiral nematic (N*), chiral antiferroelectric $N_S$*, chiral ferroelectric nematic ($N_F$*) and a chiral $N_{TBF}$*structure. The chiral pitch $p$ was evaluated via POM on cooling from the N* phase in a Grandjean-Cano (GC) wedge cell (EHC) with wedge angle $\theta = 0.68°$, minimum thickness approximately 1.6 µm and parallel rubbing along the wedge direction. Figure 2.a shows GC textures at different temperatures (full sequence in Movie S2). In the N* phase, the nonzero thickness in the thin edge promotes the π-twist state, so GC regions start with this configuration. With increasing thickness, the structure adapts to the confinement creating a series of parallel line defects perpendicular to the wedge at a distance $L$, separating GC regions of different numbers of director twists. Within the observed thickness range, only thin dislocations are observed, through which the structure twist jumps by π. Cooling into the $N_S$* phase, disclination lines shift slightly towards thicker regions, but are preserved. After the strong destabilization at the transition to the $N_F$*, the structure relaxes back into GC twist regions, now separated by edge dislocations accommodating 2$\pi$ structure changes. This behavior has been well understood,[33,34] considering that in the $N_F$* phase, the rubbed surface layer also determines the direction of polarization **P**,[43,44] limiting to structures with an even number of π-twists under parallel rubbing. Accordingly, pitch is determined by $p = 2L\tan\theta$ in the N* and $N_S$* phases and $p = L\tan\theta$ in the $N_F$* phase, respectively (Figure 2.b). While its behavior in the N* and $N_S$* phases resembles that of the parent material DIO[45], it behaves differently in the $N_F$* phase, showing a nearly constant $p$ over more than 40 degrees, while that of DIO (for similar doping characteristics) shows a pronounced linear increase on cooling.

We additionally evaluated the thicknesses (l) at which edge dislocations appear relative to the measured $p$ (Figure 2.c). While in the N* and $N_S$* phases GC lines are located as expected at the approximate thicknesses (2n+1)$p$/4 with a slight increase at the transition between both phases, in the $N_F$* they slightly shift to larger thicknesses, approximately by $p$/8.

**F7* (70:30 DIO:C1) + 1wt% S811**

N* - 99.4 °C - $N^*_S$ - 88.8 °C - $N^*_F$ - 29 °C[a] - $N^*_{TBF}$ - 20.9 °C- $SmC^{P}_{H}$*

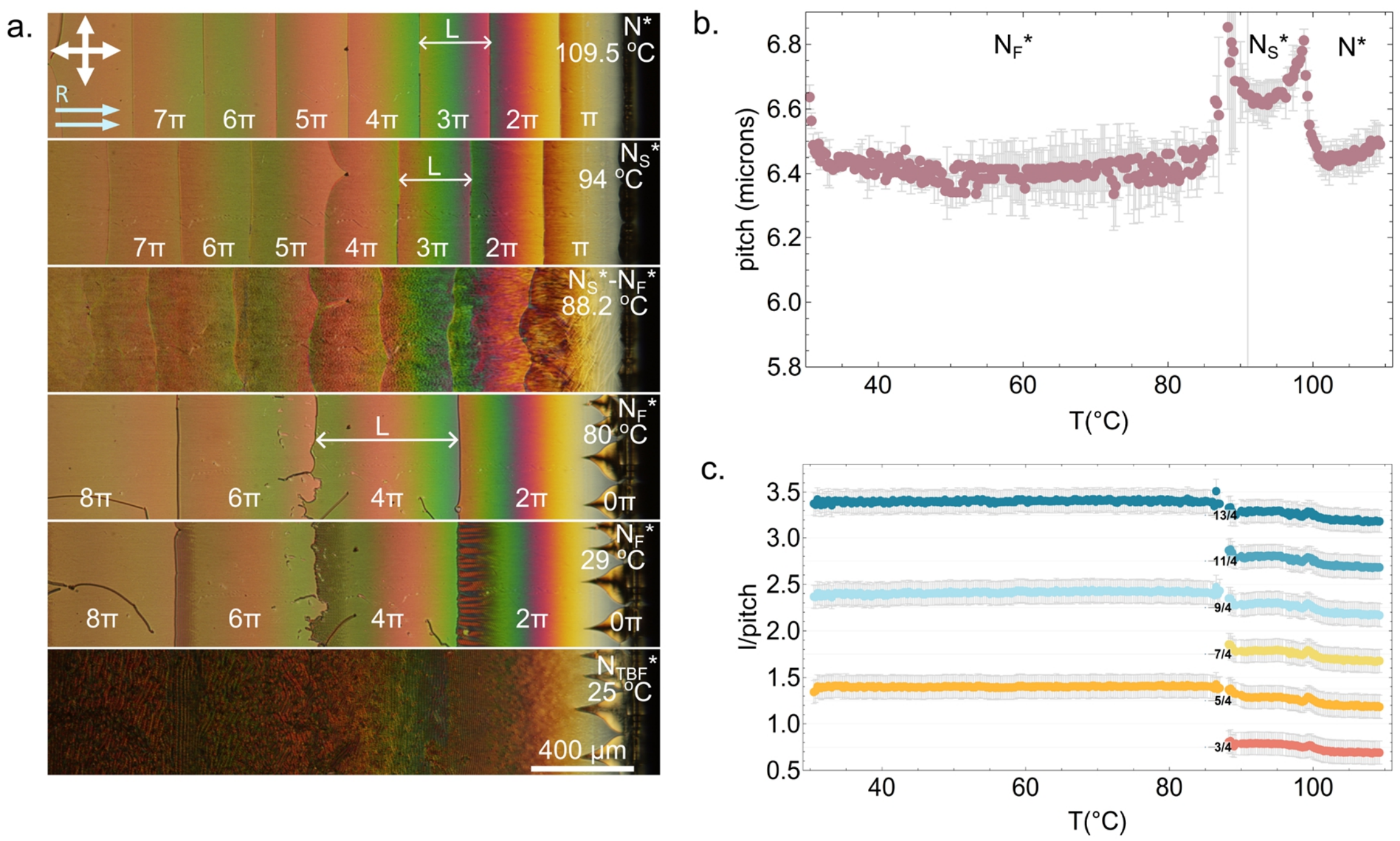

**Figure 2. Characterization of F7*** a) Temperature evolution of Grandjean-Cano texture in a wedge cell ($\theta = 0.68°$) with parallelly rubbed surfaces. The chiral twisted structure adapts to the varying thickness by creating regions of different number of director π-twist, by jumps of π (N and $N_S$) and 2π ($N_F$), separated by defect lines. Before the transition to the $N_{TBF}$ phase evident pretransitional modulated instabilities occur in the hypotwisted part of the GC regions and at the transition, a macroscopic modulated structure develops in the $N^*_{TBF}$ with period increasing with the cell thickness. b) Mesophase pitch vs temperature as determined considering $p = 2L\tan\theta$ for the N and $N_S$ phases, and $p = L\tan\theta$ for the $N_F$ phase. c) Temperature dependence of the position of the GC defect lines in terms of cell thickness (l) vs pitch.

This indicates that in the $N_F$* phase, the local structure stretches up to an apparent pitch of 8.8 μm (hypotwisted structure) before the 2π-4π dislocation line, and then contracts to half that value across it (hypertwisted structure). Such differentiated twisting conditions allow for the observation of a striking $N_F$* to $N_{TBF}$* pretransitional behavior (Figure 2.a and Movie S2), i.e., nucleating from the edge dislocation towards the thinner part, modulated instabilities appear, either 1D at a slight angle or parallel to the 2π-4π and 6π-8π GC line, or 2D as in the 4π-6π one. On entering the $N_{TBF}$* phase, the texture initially gets grainy and subsequently stabilizes into a modulated texture, with an underlying rope-like texture and overall modulation increasing with the thickness of the confinement (Figure S6). This macroscopic periodic structure, is equivalent to that reported by Nishikawa et al.[40] in an $N_{TBF}$ system, in which the chirality is extrinsically imposed by antiparallel boundary conditions. Same observation is shown by J. Karcz.[38] However, here, we will focus on pretransitional wave-like distortions observed in the intrinsically chiral $N_F$* phase.

## *2.2 Modulated instabilities in thin confinements*

In the case of untwisted states, either in the very thin regions of the wedge cell (Figure 2.a), or when material is confined in a 2.2 μm cell (*d/p*=0.32), no pretransitional modulation is observed (Figure S8). We thus turn to the next simplest twisted structure. We studied a hypotwisted 1-π structure with F7* confined in a 5.2 μm cell (*d/p*=0.8, where we take p=6.5 μm) with antiparallel rubbing (Figure S8). On slow cooling (0.5 °C/min), a few degrees above the $N_F$*-$N_{TBF}$* transition, a 1D modulated instability appears with a periodicity of approximately 6 μm and modulation perpendicular to the rubbing direction (Figure 3.a and Movie S3). On cooling, the modulation slowly evolves, i.e., the periodicity reduces, characterized by the propagation of dislocations, reaching a minimum value of around 2 μm. This results in the periodic modulation of the effective nonlinear optical coefficient $d_{eff}$, as seen in SHG-M images (Figure 3.c and Movie S4). The transition to the lower temperature phase is then marked by the appearance of an additional modulated structure, this time perpendicular to the rubbing direction of a periodicity around 3 μm, that progressively evolves into a structure (Figure 3.a-b) with periodicity equal to the cell thickness and at a slight angle with respect to surface alignment. These textural features in the $N^*_{TBF}$ are fully equivalent to those observed for the non-chiral $N_{TBF}$ under the same antiparallel surface conditions (Figure S5) and correspond to those reported by Nishikawa et al.[40] On heating, the same behavior is recovered in the $N_F$*, although at the onset of the instabilities, the periodicity appears lower than on cooling and monotonically increases with temperature before fading out (Figure 3.b and Figure S10).

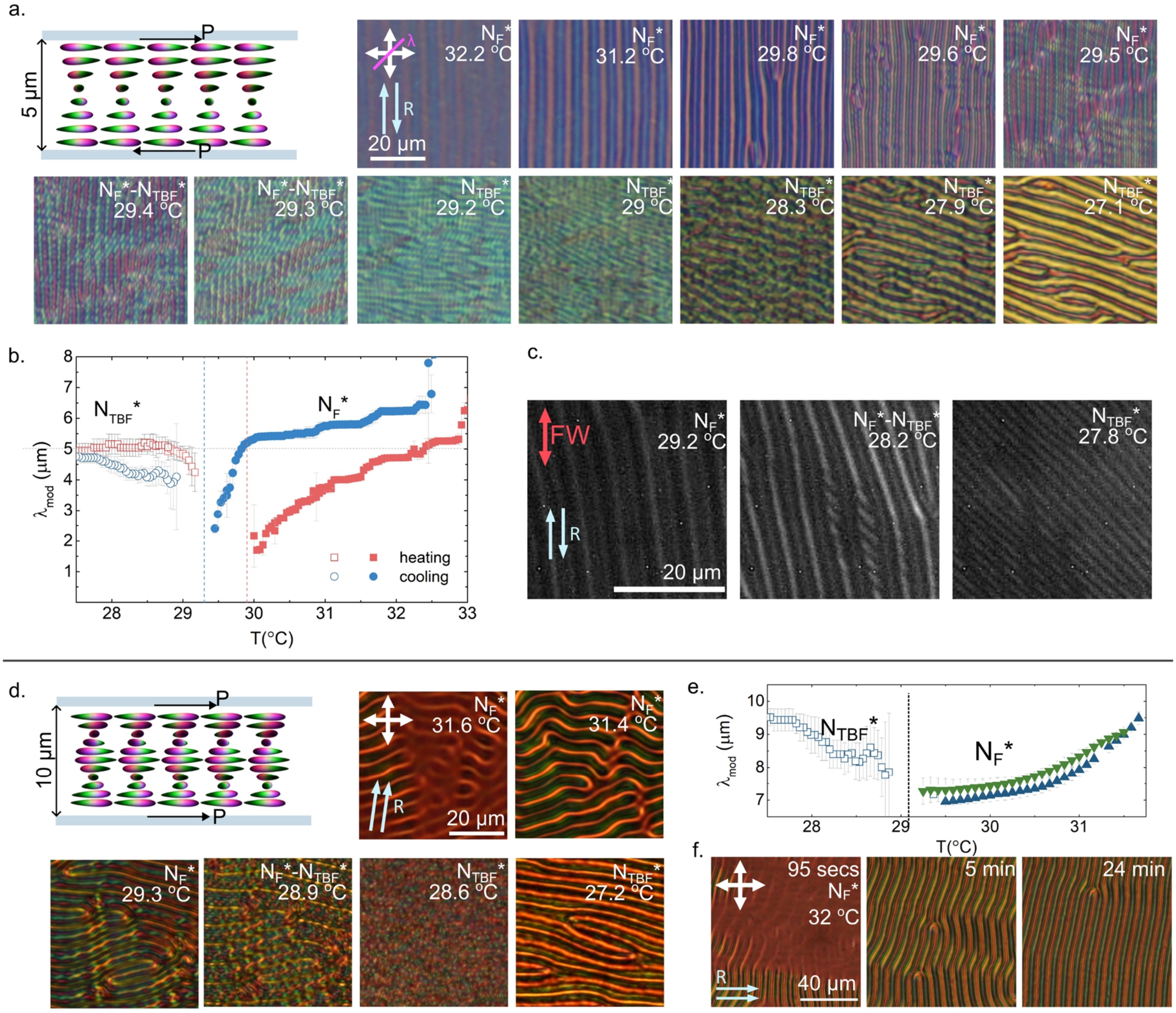

**Figure 3. Elastic instabilities in hypo-twisted π and 2π-twist structures.** a) Sketch of the π-twist equilibrium structure in a 5.4 µm antiparallel rubbed cell in the $N_F$* phase and sequence of POM images on cooling, from the onset of the instabilities in the $N_F$* phase down to the $N_{TBF}$* phase. Double-headed white arrows indicate the direction of polarizers, and light-blue arrows indicate direction of surface rubbing (R). b) Temperature evolution of the modulation of the instabilities ($N_F$*) and of the periodic macrostructure ($N_{TBF}$*) in both heating and cooling runs. c) SHG-Microscopy images of the instabilities in the $N_F$* and at the transition to the $N_{TBF}$*. Full sequence in Movie S3. Red double-headed arrow indicates direction of polarization of the fundamental beam. d) Sketch of the 2π-twist equilibrium structure in a 10.1 µm parallel rubbed cell in the $N_F$* phase and sequence of POM images on cooling, from the onset of the instabilities in the $N_F$* phase down to the $N_{TBF}$* phase. e) Temperature evolution of the modulation of the instabilities ($N_F$*) and of the periodic macrostructure ($N_{TBF}$*) on cooling. Both sets of triangles correspond to areas in which modulation appears tilted in complementary directions with respect to the rubbing direction. f) Temporal stabilization of instabilities parallel to the rubbing direction at the onset temperature.

The same analysis was conducted for a hypotwisted 2-π structure using a 10.1 µm cell ($d/p$=1.55) with both surfaces rubbed in the same direction (Figure 3.d-f). Given the confining ratio, on cooling from the N* and $N_S$* phases, the LC cell is predominantly covered by 4-π structure, with small islands of 2-π structure (Figure S11 and S12). On approaching the $N_F$* to $N_{TBF}$* transition, modulated instabilities appear in such islands, while the surrounding texture remains unaltered, allowing for a clear visualization of the importance of the elastic field in the formation of the instabilities. In this temperature interval, 2-π regions expand before the transition, but drastically in the $N_{TBF}$* phase, indicating a strong competition between the helical and the heliconical structures. Subsequent thermal treatment across the transition allows stabilization of the 2-π ground structure across the entire cell. Under the dynamic cooling conditions, modulation grows in two directions, at a slight angle from the rubbing direction, i.e., mostly perpendicularly with respect to the previous 1-π twist case. Cooling down to the onset of the instabilities and stabilizing the temperature there for a longer period (Figure 3.f) reveals a much more organized 1D morphology.

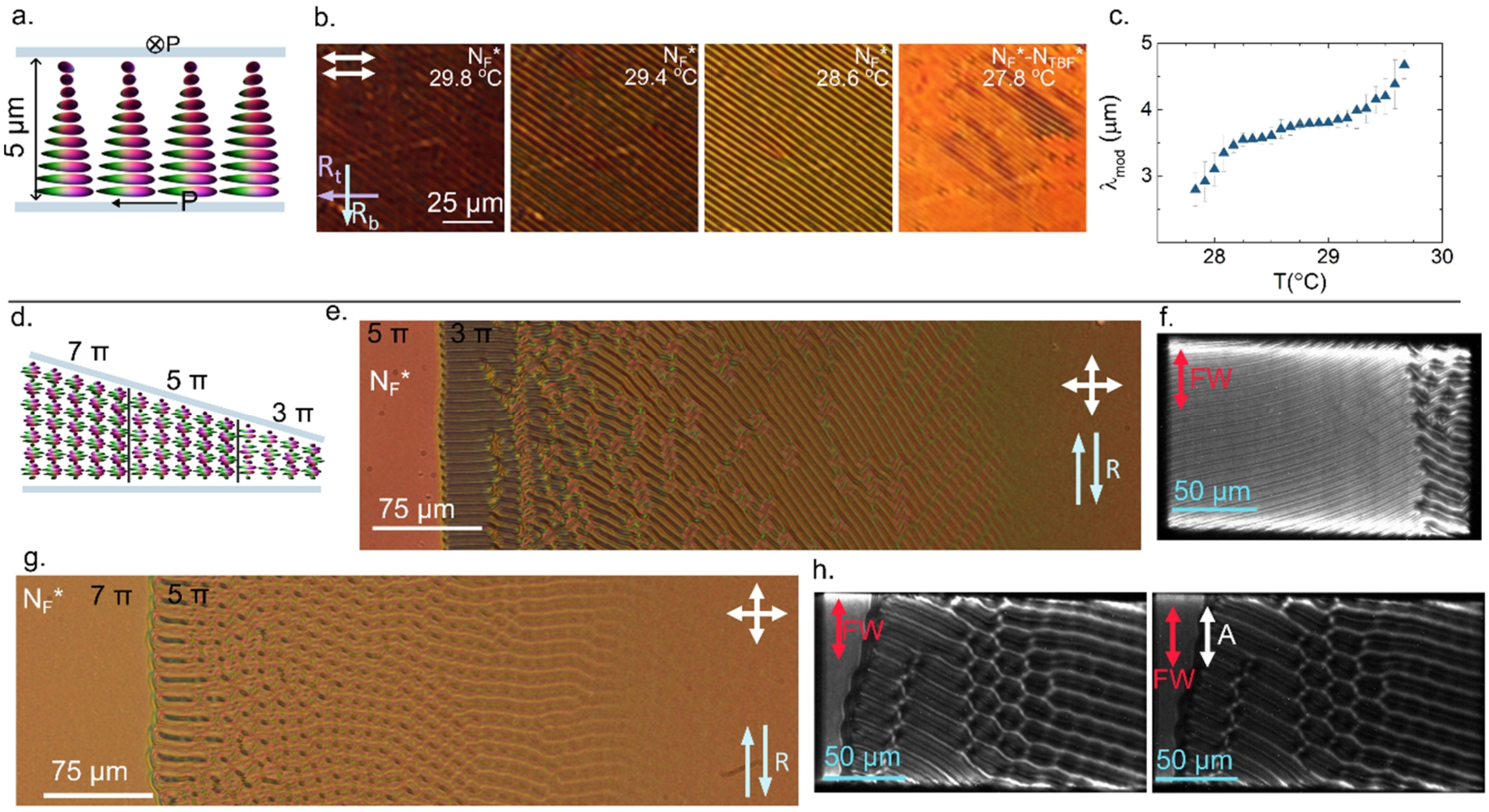

**Figure 4. Elastic instabilities in hypo-twisted π/2, 3π-twist and 5π-twist structures.** a) Sketch of the π-twist equilibrium structure in a 5 μm 90 left-handed rubbed cell in the $N_F$* phase. b) Sequence of POM images on cooling, from the onset of the instabilities in the $N_F$* phase down to the $N_{TBF}$* phase. Double-headed white arrows indicate the direction of polarizers, and light-blue arrows indicate the direction of surface rubbing (R). c) Temperature evolution of the observed modulation in the $N_F$*. d) Sketch of the custom-made wedge cell (θ=0.22°) with antiparallel rubbing perpendicularly to the wedge direction. e) POM texture in the $N_F$* phase at 29.7 °C in the boundary between the 5π and 3π-twisted structures. f) Second Harmonic Generation (SHG) microscopy image in the 3π-twisted GC region, in an equivalent area to that in (e). The red double-headed arrow represents the polarization of the incoming fundamental wave. g) POM texture in the $N_F$* phase at 29.7 °C in the boundary between the 7π and 5π-twisted structures, showing a hierarchy of 1D and 2D instabilities. h) SHG-microscopy images in the vicinity of the 7π-5π GC line, reflecting the spatial structuring of the effective second-order nonlinear susceptibility. The red double-headed arrow represents the polarization of the incoming fundamental wave, while the white arrow (A) represents the direction of the analyzer for the right panel.

The interplay between surface anchoring and modulation directions can be further explored in 1/2-π twisted structures (Figure 4.a-c), achieved in left-handed 5 μm cells (*d/p*=0.77) with buffing directions at 90 degrees (Figure S13). Under these conditions, the undulations emerge at 45 degrees, with a decreasing period on cooling.

## *2.3 Modulated instabilities in thicker confinements*

To investigate pretransitional undulations in higher twisted structures, we fabricated a home-made wedge cell with a smaller wedge angle ($\theta = 0.22°$) and antiparallel rubbing perpendicular to the wedge direction. This anchoring geometry imposes opposite orientations of **P** at the surfaces, with GC lines now separating regions with 3π, 5π, and 7π (overview and Zoom-ins in Figures S14, S15, and S16). The temperature was then held 0.5 °C below the instability onset, allowing the patterns to stabilize.

Along the GC 3π-5π defect line (Figure 4.e-f), a 1D periodic modulation appears, with stripe orientation varying with the local thickness. Near the GC disclination line, the modulation is perpendicular to the surface alignment, whereas in the thinner part of the 3π twisted region, the stripes rotate to about -45 degrees. At the same time, the periodicity of the modulation slightly reduces from approximately 6.6 to 5 μm, i.e., from values $\sim p$ to $\sim 0.8p$. It should be noted that other regions of the cell display additional periodic morphologies with increased complexity. In the thicker region, around the GC 5π-7π defect line (Figure 4.g-h), different morphologies can be observed, from 1D modulations by the GC line, to 2D square and 2D hexagonal patterns with periodicities ranging between 8 and 10 μm depending on the local thickness. All these structures are clearly resolved in SHG-M images, which show a pronounced spatial modulation of the SHG signal that closely follows the periodic morphology of the undulation patterns, with the signal being maximized when the fundamental wave polarization is parallel to the surface alignment direction (Figures S15 and S16). The modulation of both the director field and the SHG intensity indicates a periodic variation of the local polarization direction. This shows that the elastic instability produces a spatially modulated polarization field, effectively forming a polarization wave driven by the director deformation and highlights the potential of these systems as self-organized periodic nonlinear optical media.

## *2.4 Modelling of instabilities*

The above described observations differ significantly from recently reported phenomena in related systems. Particularly, Nishikawa et al. stabilized the heliconical polar structure of the HEC mesophase in a non-chiral system between antiparallelly rubbed substrates, imposing an extrinsically twisted configuration.[40] They observed that the coexistence of extrinsic (helical) and intrinsic (heliconical) twists leads to a triple chiral state, observed in POM as a striped texture, with stripes oriented perpendicularly to the rubbing direction. This structure occurs in the low temperature HEC phase at the transition from the $N_F$ phase, similarly to the observations here down in the $N_{TBF}$*, with the difference that in the present case, the helical twist is also intrinsic. In observations more closely related to ours, Zhou et al. reported the formation of a 2D grid-like pattern in the $N^*_F$ phase of chiral systems based on RM734 and homologues.[46] However, the authors identified a critical thickness (dependent on the helical pitch) above which these structures appear and modelled their emergence using a free energy functional including elastic, compression and polar contributions. Their model returns a critical pitch ($2\pi/q_c$) and the condition of a critical thickness $d_c = 2\pi N/q_c$ which prevents the formation of the modulated structure in samples thinner than the material helical pitch.[46]

Strikingly, in the case of our observations, although a 2D-grid-like pattern is observed in the thicker parts of the sample, 1D modulations are reproducibly obtained in thin confinements, even in thicknesses comparable to or smaller than the helical pitch, always in hypotwisted confinement. This behavior is similar to that reported for twist-bend nematic dimers and is driven by elastic instabilities.[25,27] For the parent non-chiral F7 material considered here, we recently reported a marked softening of the bend elastic constant $K_3$ on approaching the $N_F$ to $N_{TBF}$ phase.[47] While Figure 5.a shows the reported temperature dependence of the $K_3/K_2$ ratio, $K_2$ is shown to remain practically constant prior to the transition, with the pronounced reduction being driven by the softening of $K_3$. Here, we investigate the emergence of instabilities by determining stability diagrams in terms of confinement thickness and the elastic ratio $K_{3n}=K_3/K_2$. The director field is parametrized in terms of two angular variables, $\theta(\mathbf{r})$ in the x-y plane and $\alpha(\mathbf{r})$ in the out of plane direction, $\mathbf{n} = (\cos\alpha(\mathbf{r})\cos\theta(\mathbf{r}), \cos\alpha(\mathbf{r})\sin\theta(\mathbf{r}), \sin\alpha(\mathbf{r}))$. Here the base state depends only on the coordinate z along the cell thickness. Then the stability is analyzed with respect to small periodic perturbations of the type $\alpha(\mathbf{r}) = \alpha_0(z) + \delta\alpha(z)e^{i\boldsymbol{q}\cdot\boldsymbol{r}_\perp} + c.c.$ and $\theta(\mathbf{r}) = \theta_0(z) + \delta\theta(z)e^{i\boldsymbol{q}\cdot\boldsymbol{r}_\perp} + c.c.$, with the wavevector $q = (k_x, k_y)$ and $\boldsymbol{r}_\perp = (x, y)$. To account for the different N-twisted structures, we take $\theta_0(z) = N\pi z/d$ and $\alpha_0(z) = 0$.

We first establish the equilibrium configuration by numerical minimization of the Frank-Oseen free energy

$$F = \int \tfrac{1}{2}K_1(\nabla\cdot\mathbf{n})^2 + \tfrac{1}{2}K_2(\mathbf{n}\cdot(\boldsymbol{\nabla}\times\mathbf{n}) + q_0)^2 + \tfrac{1}{2}K_3(\mathbf{n}\times(\boldsymbol{\nabla}\times\mathbf{n}))^2\, dV - \int \frac{1}{2}W(\mathbf{n}\cdot\mathbf{n}_s)^2 dS\,,$$

where $K_i$ are the elastic constants, $q_0 = 2\pi/p$ describes the helical pitch periodicity. In the last term $\mathbf{n}_s$ is the preferred direction of **n** at the surface of the cell, which is taken parallel to the substrates, and *W* is the anchoring strenght. It is important to note here that, despite being a polar phase, for now, we will be just considering elastic terms. The stability analysis is performed numerically by calculating for each pair of parameters *d/p* and $K_3/K_2$, the lowest nontrivial eigenvalue and corresponding wavevector **q**. The onset of the instability is then identified when the minimum eigenvalue becomes negative at a finite **q**, marking the transition of the undeformed state towards the modulated instability. The calculations were performed considering $W = 1\cdot10^{-5}\ \mathrm{Jm}^{-2}$ and $K_{1n}=K_1/K_2=100$, the latter reflecting the electrostatic cost ($\rho_b = -P_0\nabla\cdot\mathbf{n}$) of splay deformations of the polar system. To assess this effect, additional calculations with $K_{1n}=2.5$ were also performed.

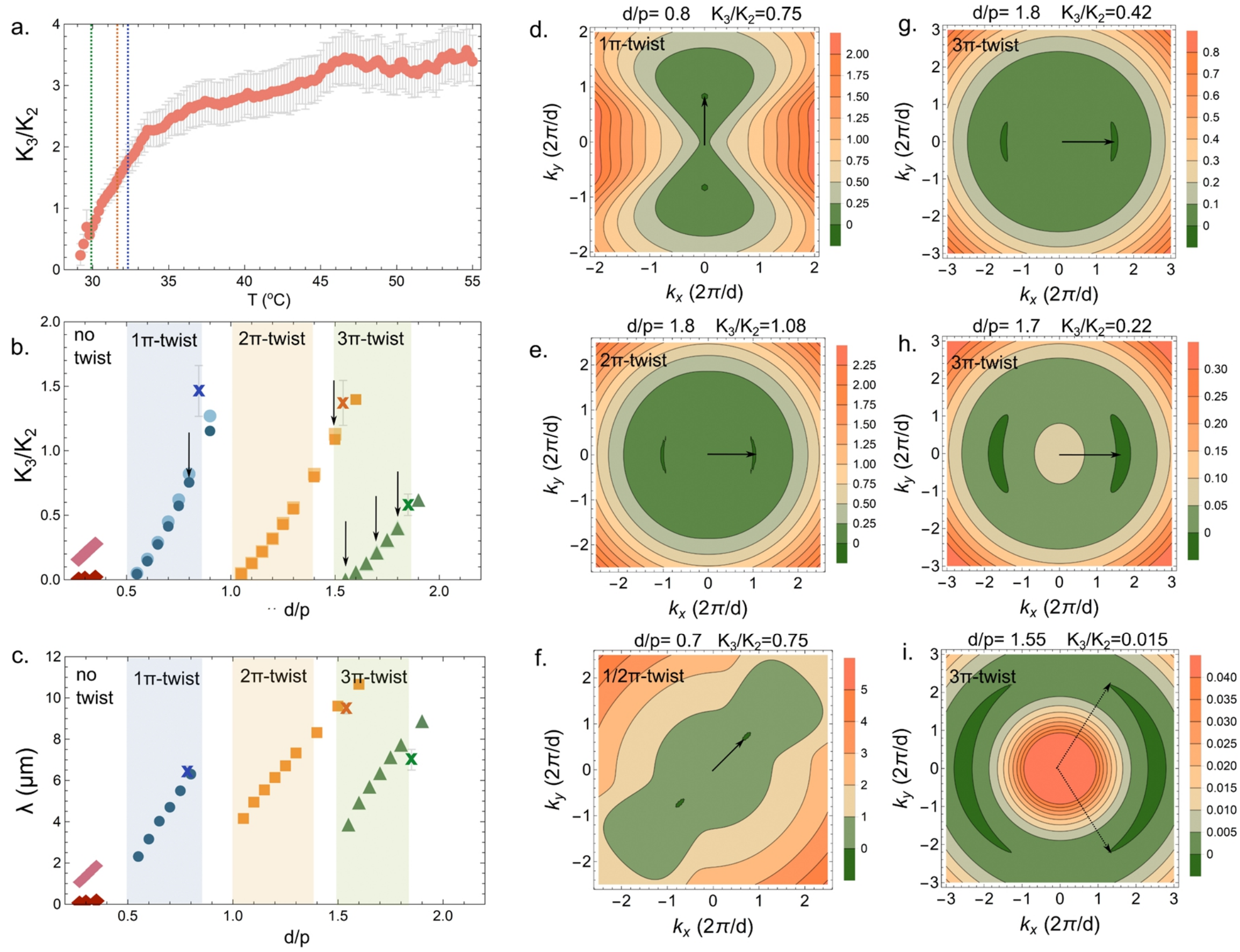

**Figure 5. Numerical Predictions vs. Experimental Observations of Twist-Induced Elastic Instabilities.** a) Temperature dependence of the ratio between the bend and twist elastic constants $K_3/K_2$ as reported in [46] used to compare with the model results. b) Critical $K_{3n}=K_3/K_2$ as a function of confinement ratio d/p for 0π-, 1π-, 2π- and 3π- twisted ground structures as determined from linear stability analysis. Lighter symbols correspond to $K_{1n}$=2.5, while darker ones correspond to $K_{1n}$=100. Crosses mark for the experimental *d/p* investigations, the corresponding $K_{3n}$ value as obtained from (a). c) Calculated instabilities modulation for the 0π-, 1π-, 2π-, and 3π- twisted ground structures are compared with the experimentally observed ones at the onset of instabilities for the different confinements. (d-i) Two-dimensional maps of the lowest calculated eigenvalues as a function of the instability wavevector ($k_x$,$k_y$) for different confinement ratios and twisted ground structures Vertical arrows in b) correspond to the instability points shown in (d-i). For 1π- (d), 2π-(e) and 1/2π-(f) structures, the anisotropic minima select stripe-like modulations with modulation perpendicular, parallel, and at 45 degrees with respect to surface alignment direction (along x), respectively. (g-i) For the 3π-twist structure, the dispersion shows a strong dependency on the confinement ratio, showing a shallow minimum allowing for multiple accessible modulation directions as experimentally observed.

Stability diagrams for the structures $\theta_0(z) = N\pi z/d$ with N=0, 1, 2, and 3 are shown in Figure 5.b within the hypotwisted *d/p* bands. For N=0, it is found that $K_{1n}$=100 suppresses the instability for the untwisted state, in agreement with observations here and in clear contrast to the findings in twist-bend dimers.[27] For the twisted structures, the effect of $K_{1n}$ in the stability diagram is minimal (Figure 5). The model reveals an inverse relationship between the required $K_{3n}$ and the deviation of the imposed twist from the intrinsic cholesteric pitch $q_0$: As the frustration of the natural pitch increases, instabilities are triggered at larger bend elasticity. No unstable solutions are found for the hypertwisted structures (i.e. *d/p* < 0.5, 1, and 1.5). These results are compared with experimental observations for onset conditions and modulation periodicities at *d/p* = 0.8, 1.55, and 1.8, where the corresponding $K_{3n}$ values are extracted from Figure 5.a at the observed onset temperature. Experimental values are indicated by crosses in Figure 5.b&c.

Further insight into the nature of the emerging patterns is obtained by analyzing the spectral landscape in the ($k_x$,$k_y$) plane. Representative plots for twisted states corresponding to the structures with 1π-, 2π-, 1/2π-, and 3π- twist across the cell are shown in Figure 5.d-i. In the first three cases, two symmetric narrow minima appear, indicating that the instability selects a preferred orientation and therefore favors stripe-like modulations. The orientation depends on the initial twist structure, perpendicular to the surface direction for 1π, parallel in the case of 2π, and at 45 degrees for 1/2π, in line with our experimental observations. The calculated wavelengths are shown in Figure 5.c. To reproduce the experimental behavior in the wedge cell for the 3π-structure, the same analysis is performed for different confinement ratios *d/p*. Reducing d/p results reveal a progressive increase in the number of accessible modes with different orientations, together with a slight increase in the critical *k* value, evidenced by the shallow minimum in the spectra. This matches experimental observations shown in Figure 4, where instabilities first emerge from the GC line, and expand on cooling towards the thinner regions while the angle varies and the period decreases. The same analysis is extended for the 5π (Figure S17), revealing a nearly isotropic minimum at a finite wavevector, corresponding to modes with different orientations but similar magnitudes. This allows for different combinations of critical modes, including single-mode stripes, orthogonal mode pairs, forming square lattices, and triplets, that can give rise to hexagonal lattices, consistent with the hierarchy of patterns observed experimentally.

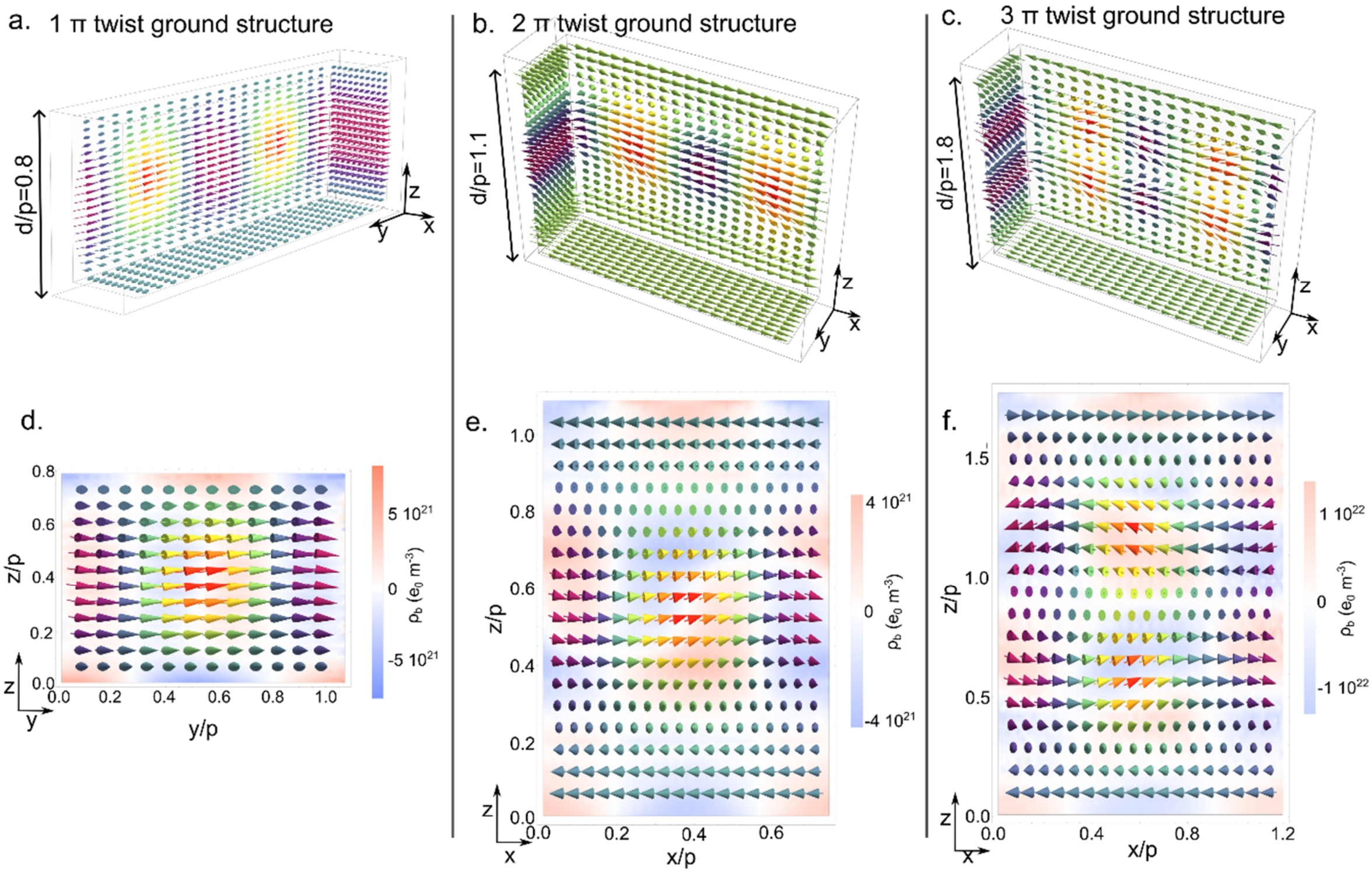

**Figure 6. Director fields and bound charge density.** a-c) Director/polarization fields obtained by taking the critical instability mode as an initial condition for free energy minimization for 1π-, 2π-, and 3π- twist structures, where x is along the cell surface alignment direction and z is along the cell thickness. The colour of the arrows indicates differences in the z component of the local director. (d-f) Projected director/polarization field for the structures in (a-c) and bound charge density as calculated from $\rho_b = -\nabla \cdot \boldsymbol{P}$ considering $\mathbf{P} = P_0 \mathbf{n}$ and $P_0 = 5\ \mu\mathrm{C/cm^2}$.

As the system is polar, these director distortions imply periodic modulation patterns of the spontaneous polarization **P**. These variations generate bound charges, according to $\rho_b = -\nabla \cdot \boldsymbol{P}$, which can introduce additional electrostatic cost. To evaluate this effect, we reconstruct the three-dimensional director configurations associated with the unstable modes and calculate the corresponding bound charge density, assuming $\mathbf{P} = P_0 \mathbf{n}$. The director fields are obtained by taking the critical instability mode as an initial condition for free energy minimization. The resulting configurations are shown in Figure 6. From the calculated director field and considering $P_0 = 5\ \mu\mathrm{C\ cm^{-2}}$ we then determine the spatial distribution of bound charges. Considering a typical ion density value $\rho_0 = 1 \cdot 10^{23}\ \mathrm{ions\ m^{-3}}$ value for the system free charges, calculated bound charge density is significantly smaller, implying that mobile ions can efficiently screen polarization charges. We can further consider the Debye screening length $\lambda_D = \sqrt{(\varepsilon \varepsilon_0 k_B T)/(2\rho_0 e^2)}$ and the characteristic length associated with bound charges that can be calculated for a periodic distortion $\xi_b = \sqrt{(\varepsilon \varepsilon_0 k_B T)/(P_0 k A_0)}$, with $A_0$ the amplitude of the modulation. Considering the obtained modulations and amplitudes (Figure S18), the coefficient $\beta = \xi_b / \lambda_D$ is larger than 1 for all cases, indicating that screening in these conditions is efficient, keeping electrostatic energy small and allowing for the development of the instabilities as observed.

## 3. Outlook

The good correspondence between the model and our experimental observations shows that the observed instabilities arise purely from elastic energy considerations. In this respect, this is analogous to what happens in twist-bend dimers, where the elastic field, effectively played by the elastic frustration and the low energetic cost of bend deformations, directly generates modulated director fields without the need for external electric or magnetic fields.[25,27] However, in the present case, the coupling between director distortions and spontaneous polarization implies that elastic instabilities in ferroelectric nematic fluids naturally generate periodic polarization structures, directly translating into spatial structuring of the effective second-order nonlinear susceptibility. This provides a direct route for creating polarization patterns through purely mechanical or geometric control of the director configuration. Temperature stability of the patterns could be improved by selective doping of FNLC materials with bend dimers, as has been proven efficient in the past for other systems.[48,49] This ability to generate polarization-modulated states through purely elastic and geometric control offers a powerful approach toward reconfigurable photonic materials.

More broadly, these results highlight FNLCs as a promising soft-matter platform for exploring polar topological structures. The coexistence of chirality, polarity, and strong elastic anisotropy allows the emergence of polarization textures conceptually related to those topological formation states (from vortices to hopfions) observed in solid ferroelectrics.[50–55] However, the soft matter nature enables continuous tuning of these structures through confinement, temperature, and elastic parameters in a more accessible way, offering a powerful tool for probing fundamental aspects of polarization topology in condensed matter.

## 4. Experimental Section

### Materials

The synthesis of the liquid-crystalline material DIO (2,3',4',5'-tetrafluoro-[1,1'-biphenyl]-4-yl 2,6-difluoro-4-(5-propyl-1,3-dioxan-2-yl)benzoate) has been carried out according to the description given in reference [29]. The molecular structure, together with the phase sequence are presented in Figure S1. Description of the synthesis of the liquid-crystalline material C1 can be found in reference [41]. The molecular structure and the phase sequence are also given in Figure S1.

F7* mixture was prepared by mixing F7 binary mixture (70:30 DIO:C1) and the commercial chiral dopant S811 (Sigma-Aldrich), with the later at 1wt%. F7 and S811 were weighted separately, adding the required volume of a S811 solution at 1mg/ml into F7. Additional chloroform was then added for full dissolution, and then the mixture was sonicated for 5 minutes. After solvent evaporation, the mixture was let to dry for 2 hours at 70 °C. Due to the 1,3-dioxane unit in both materials, the samples were always maintained below 120 °C to avoid changes in the molecular structure.

In this work, we employed both commercial and custom-made liquid crystal cells. Commercial cells were acquired from EHC, with cell gaps ranging from 2 to 25 µm (KSRP D-type) and combining parallel or antiparallel rubbing (polyimide LX-1400 from Hitachi-Kasei) as specified in the results section. EHC wedge cells with $tan\theta = 0.0115$ according to specifications, which were subsequently calibrated. The temperature dependence of the wedge cells' slope was calibrated before filling, detecting only a slight change above 85 °C, accounting for a max 1.2% change at 110 °C, which was restored after cooling. Homemade cells were built with soda-lime glass 1x2cm coated with indium-tin-oxide treated with a 30% solution of polyimide SUNEVER 5291 (Nissan) film, and rubbed to achieve orientational in-plane anchoring (planar alignment) of the LC. To build the wedge cell, glass spacers were employed.

### High-resolution AC calorimetry

High-resolution AC calorimetric measurements were carried out using a home-built experimental setup. A comprehensive description of the method is provided in Ref. [56]. The system offers thermal stability better than 0.1 mK and allows slow cooling and heating scans. The usual AC mode of operation precisely yields the heat capacity dependence on temperature for a second-order (continuous) phase transition. In the case of a first-order (discontinuous) transition, the phase coexistence region is indicated by an anomalous behavior of the phase shift between the applied AC power and the temperature oscillations of the sample. The latent heat can be determined by performing additional relaxation runs. A sample with a mass of 30 mg was loaded into a silver cell, to which a heater and a thermistor were attached on each side. Prior to measurements, the sample was heated to the isotropic phase at 110 °C. A cooling run was then carried out at a rate of 350 mK/h, through the temperature range where the N-$N_S$ and $N_S$-$N_F$ transition was expected. Subsequently, the sample was cooled at a faster rate down to 42 °C, where measurements commenced at a cooling rate of 400 mK/h.

### X-Ray diffraction measurements

Two dimensional diffraction diagrams were recorded using a Stoe Stadivari goniometer equipped with a Genix3D microfocus generator (Xenocs) and a Dectris Pilatus 100K detector using Cu-K$\alpha$ radiation ($\lambda$ = 1.5418 Å). A nitrogen-gas Cryostream controller (Oxford Cryosystems) allowed for a temperature control of about 0.1 °C. The materials were introduced at 90 °C in glass capillary tubes of 0.6 mm of diameter, which was continuously rotated during pattern adquisition.

### Polarising Optical Microscopy

POM experiments were performed in an Optiphot-2 POL Nikon microscope. Images and videos were recorded with a Canon EOS M200 camera. The sample was held in a heating stage (Instec TS62) together with a temperature controller (STC200, Instec).

### Second Harmonic Generation Microscopy

Second Harmonic Generation microscopy (SHG-M) was performed using a custom-built sample-scanning microscope with an Erbium-doped fiber laser (C-Fibre A 780, MenloSystems, 785 nm, 95 fs pulses at a 100 MHz repetition rate) as the laser source. Average power is adjusted using an ND filter depending on the experimental needs. A combination of galvo mirrors and a long-working distance objective (Nikon CFI T Plan SLWD, NA 0.3) is used to scan the focused beam in the sample plane. The scanning frequencies are much higher (a few 100 Hz) than the imaging frame rate (a few Hz). A detailed description of the setup is available in reference [57]. A long-working distance 20× objective (Nikon CFI T Plan SLWD, NA 0.3) collects the light coming from the sample and a set of 700 nm short-pass and 400 nm band-pass filters eliminates the fundamental IR light and any possible fluorescence signal. The final SHG-M images are acquired with a high-performance CMOS camera (Grasshopper 3, Teledyne Flir) with a typical integration time of 250 ms, varying dimensions in pixels, and a resolution of 0.285 µm/pixel, or in case of fast scanning 1.12 µm/pixel.

## Acknowledgements

The authors thank Janja Milivojević for the fabrication of the custom-made liquid crystal cells. A.S., L.C., P.M.R., M.L., N.O., A.M. and N.S. acknowledge the Slovenian Research and Innovation Agency (ARIS, grant no. P1-0192, P1-0125 and J1-50004, BI-VB/25-27-011) for financial support. C.J.G, J.H. and R.J.M acknowledge funding from UKRI via a Future Leaders Fellowship, grant no. MR/W006391/1. A.E. and J.M.-P. acknowledge funding from the Basque Government Project IT1979-26 and from project PID2023-150255NB-I00 from MCIU/AEI/10.13039/ 5011000-11033/FEDER, UE.

## Author Contribution

N.S designed and coordinated the work. C.J.G., J.H. and R.J.M developed the non-chiral mixture with room temperature $N_{TBF}$ phase. N.S. prepared chiral materials and conducted POM observations and analysis. A.S conducted additional POM investigations. A. S. and P.M.R. carried out SHG experiments. N.O. developed and upgraded the SHG-M setup. M.L. performed calorimetry experiments. A.E. and J.M-P. carried out X-ray, spontaneous polarization and optical diffraction measurements. A.M. developed the numerical code and N.S. implemented it. N.S. prepared the initial manuscript and all the authors read it and made contributions to the final version.

# Supporting Information

## Elasticity-Driven Periodic Polarization Patterns in Confined Chiral Ferroelectric Nematic Fluid

Anej Sterle[1,2], Peter Medle-Rupnik[1,2], Luka Cmok[1,2,3], Aitor Erkoreka,[4] Marta Lavrič[1], Natan Osterman[1,2], Calum J. Gibb[5], J. Hobbs[6], Josu Martinez-Perdiguero,[4] Richard J. Mandle[5,6], Alenka Mertelj[1], Nerea Sebastián[1*]

[1] Jožef Stefan Institute, Ljubljana, Slovenia
[2] Faculty of Mathematics and Physics, University of Ljubljana, Ljubljana, Slovenia
[3] CENN Nanocenter, Ljubljana, Slovenia
[4] Department of Physics, Faculty of Science and Technology, University of the Basque Country UPV/EHU, Bilbao, Spain
[5] School of Chemistry, University of Leeds, Leeds, UK
[6] School of Physics and Astronomy, University of Leeds, Leeds, UK

*corresponding author: nerea.sebastian@ijs.si

## **Supporting F7 mixture characterization**

DIO: Cr ( $N_F$ 65 °C) - $N_S$ - 84.7 °C - N - 174 °C - Iso

C1: Cr ($SmC^H{}_P$ 90.1 °C) - $SmA_F$ - 129.7 °C - SmA - 154.3 °C - N - 225.6 °C - Iso

Supporting Figure 1. Molecular structure and phase sequence of DIO [1] and C1 [2].

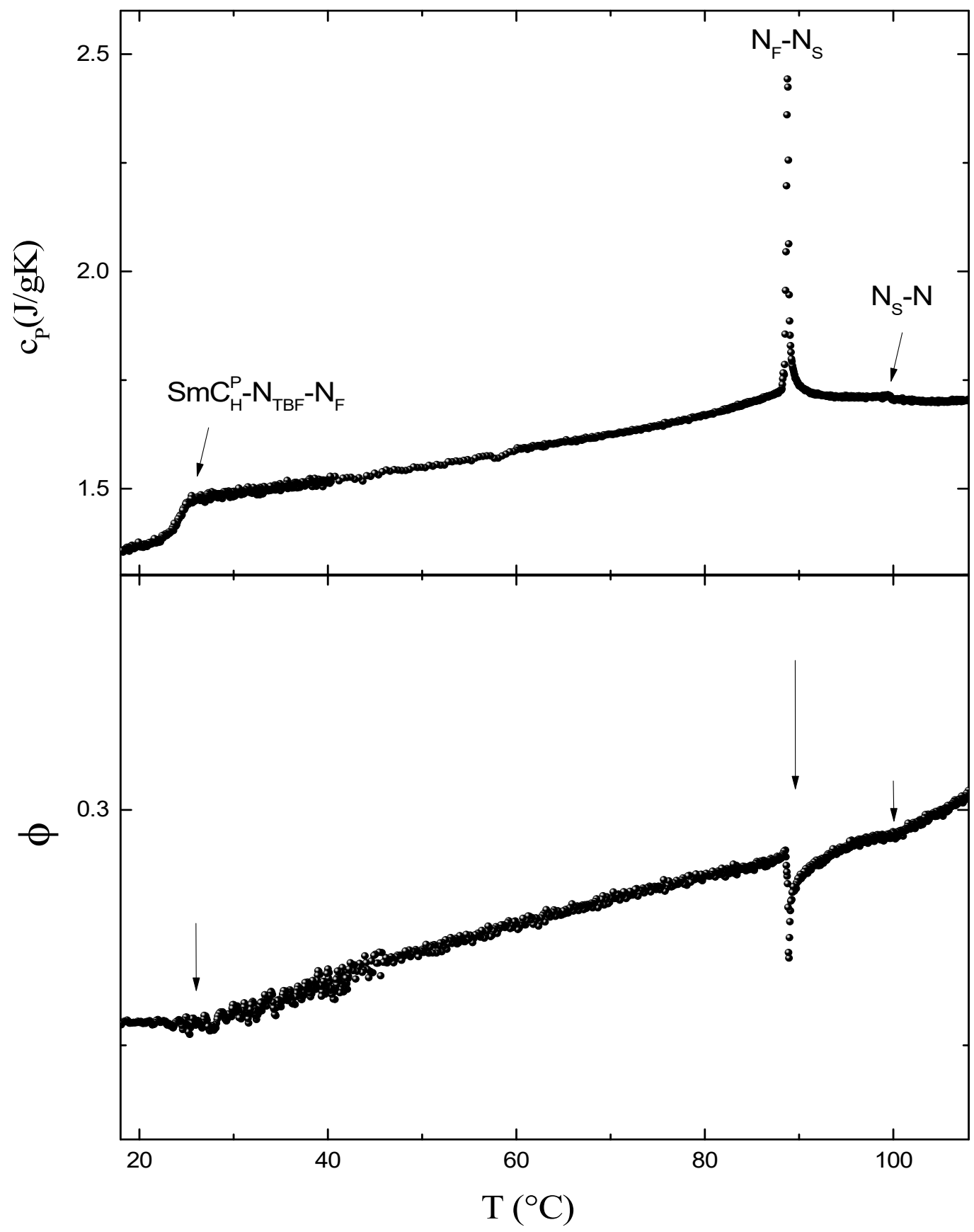

Supporting Figure 2. (top) The temperature profile of the heat capacity, $c_P(T)$, for the F7 is shown upon cooling. The data were obtained by high resolution AC calorimetry. (bottom) Temperature dependence of the corresponding phase shift showing negative peaks, indicating the second-order character of the transitions.

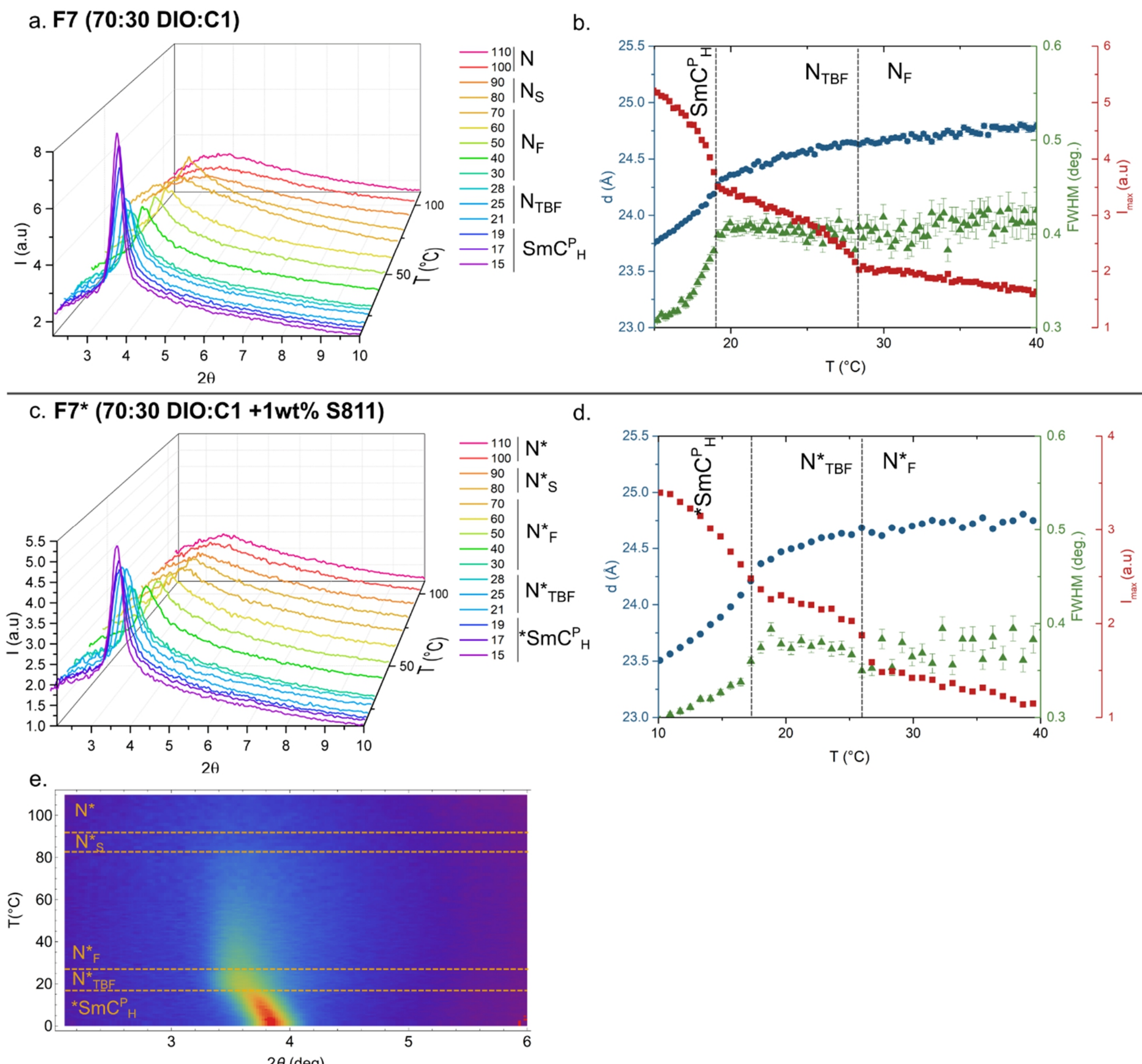

Supporting Figure 3. X-Ray experiments in F7 (a,b) and F7* samples (c,d,e). a&c) Azimuthally integrated X-ray diffractograms from 2D wide-angle non-aligned diffraction patterns at selected temperatures from the dataset in Fig. 1f. The initial broad signal characteristic of the N phase, progressively narrows in the $N_F$ phase, showing pretransitional smectic fluctuations already throughout the $N_F$ phase. This behaviour has previously been reported for other materials exhibiting the ferroelectric heliconical phase [3-5]. b&d) Temperature dependence of the local correlations periodicity ($N_F$ and $N_{TBF}$) and layer spacing ($SmC^P_H$), amplitude of the scattering peak and FWHM extracted from the data presented in Fig.1.f & Figure S3.e . Peak intensity strongly marks phase transitions. At the $N_F$-$N_{TBF}$ transition, a faster gradual reduction of d is observed while the FWHM plateaus. This is compatible with the growth of the twist-bend modulation without an increase in the correlation length. At the $N_{TBF}$-$SmC^P_H$ transition, a fast reduction of *d* is observed as well as a strong narrowing of the signal compatible with the transition to a higher-ordered tilted smectic phase. (e) Temperature evolution of the X-ray diffractograms obtained by integration of the 2D Broad-angle XRD for an unaligned F7* sample.

## SHG and polarization

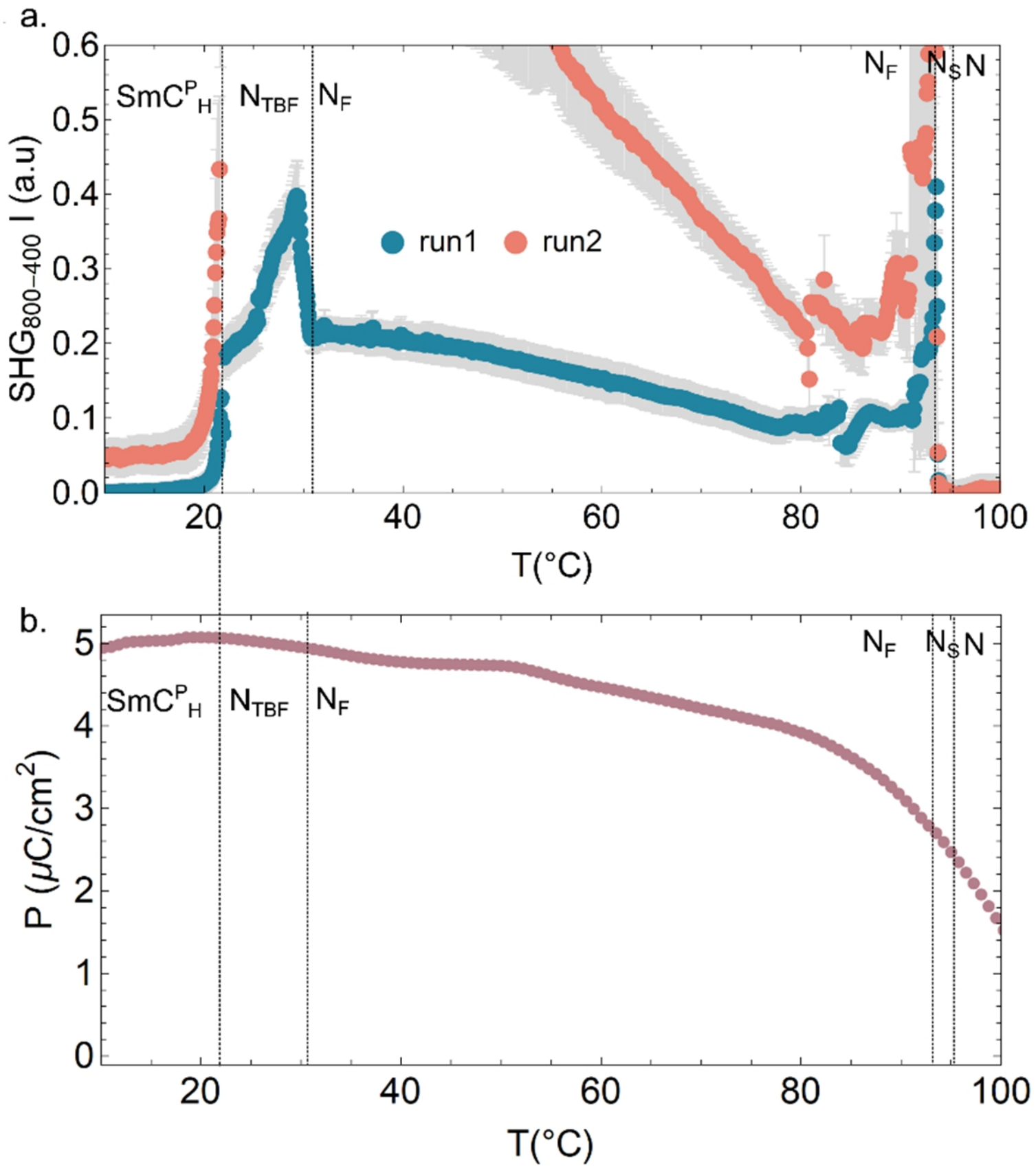

Supporting Figure 4. a) Temperature dependence of SHG signal for F7 recorded in a 5 μm parallel rubbed cell. Both runs were performed on cooling, with the fundamental beam polarization along the cell rubbing direction and no analyzer. For run2 the detector exposure time and gain were increased in order to raise the observed SHG signal in the $SmC^P_H$ phase above the noise level. b) Temperature dependence of spontaneous polarization in the $N_F$, $N_{TBF}$ and $SmC^P_H$ phases. Values measured in the temperature range of the N and $N_S$ phase correspond to induced polarization due to the high applied voltages required for polarization saturation in the $N_F$, $N_{TBF}$ and $SmC^P_H$ phases. Polarization values are obtained via the application of a triangular wave voltage of 12 Hz and 25 Vpp amplitude in a 5 μm thick untreated cell, via the integration of the current peak measured with a 1 kOhm resistor.

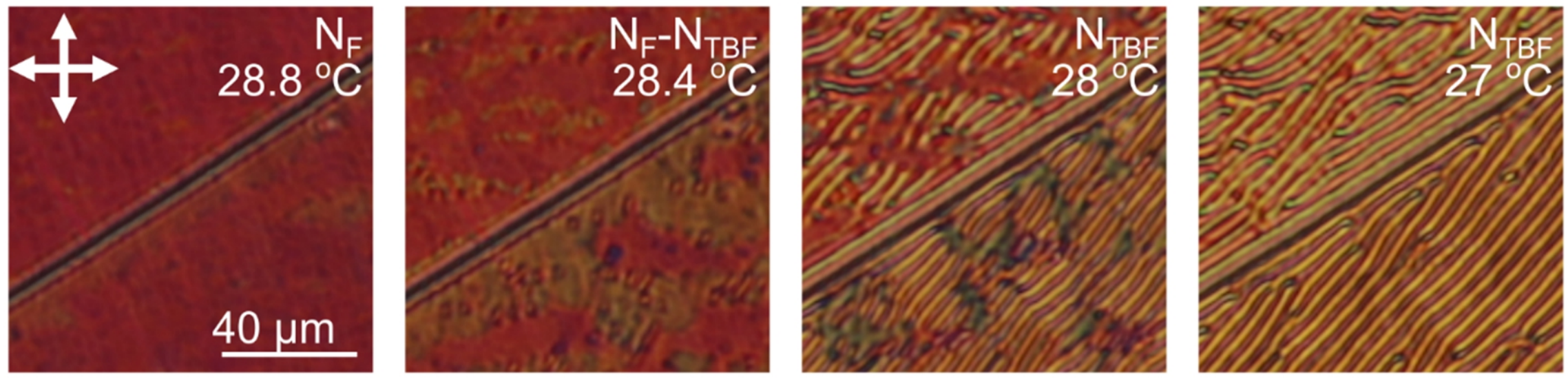

Supporting Figure 5. Polarizing optical microscopy images of F7 sample confined in a 5 μm antiparallel rubbed cell around the $N_F$-$N_{TBF}$ transition showing the formation of hierarchical 3D twist in the $N_{TBF}$ phase as described by J. Karcz et al [3] and Nishikawa et al. [4]

## Supporting F7* mixture characterization

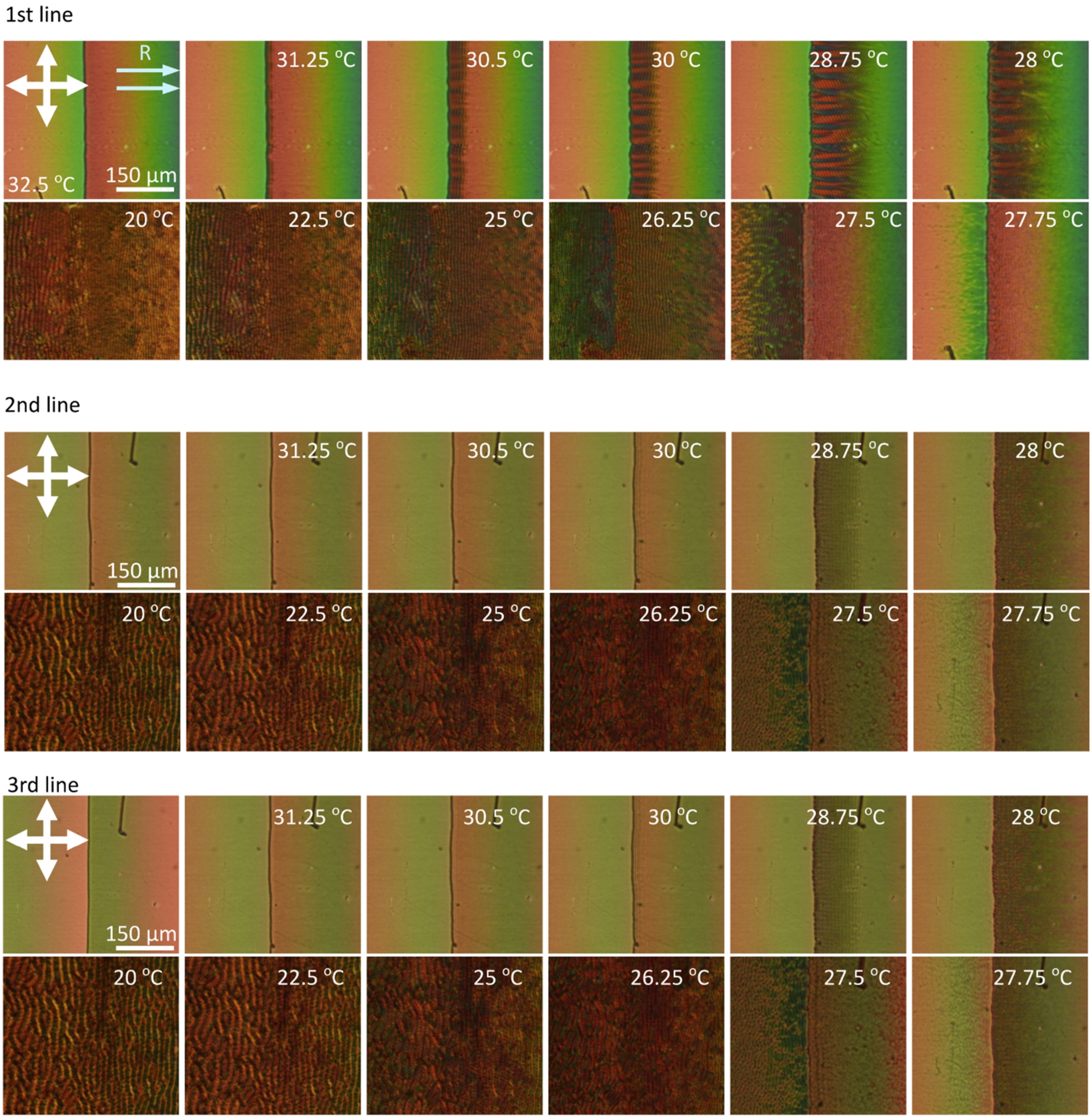

Supporting Figure 6. Zoom-ins around the $1^{st}$, $2^{nd}$ and $3^{rd}$ GC-lines in a wedge cell (EHC) with wedge angle $\theta = 0.68°$ and parallel rubbing along the wedge. First row of each panel corresponds to the evolution of the instabilities in the $N_F$* phase, while the second row corresponds to the transition to the $N_{TBF}$* phase and the formation of a macroscopic modulated structure.

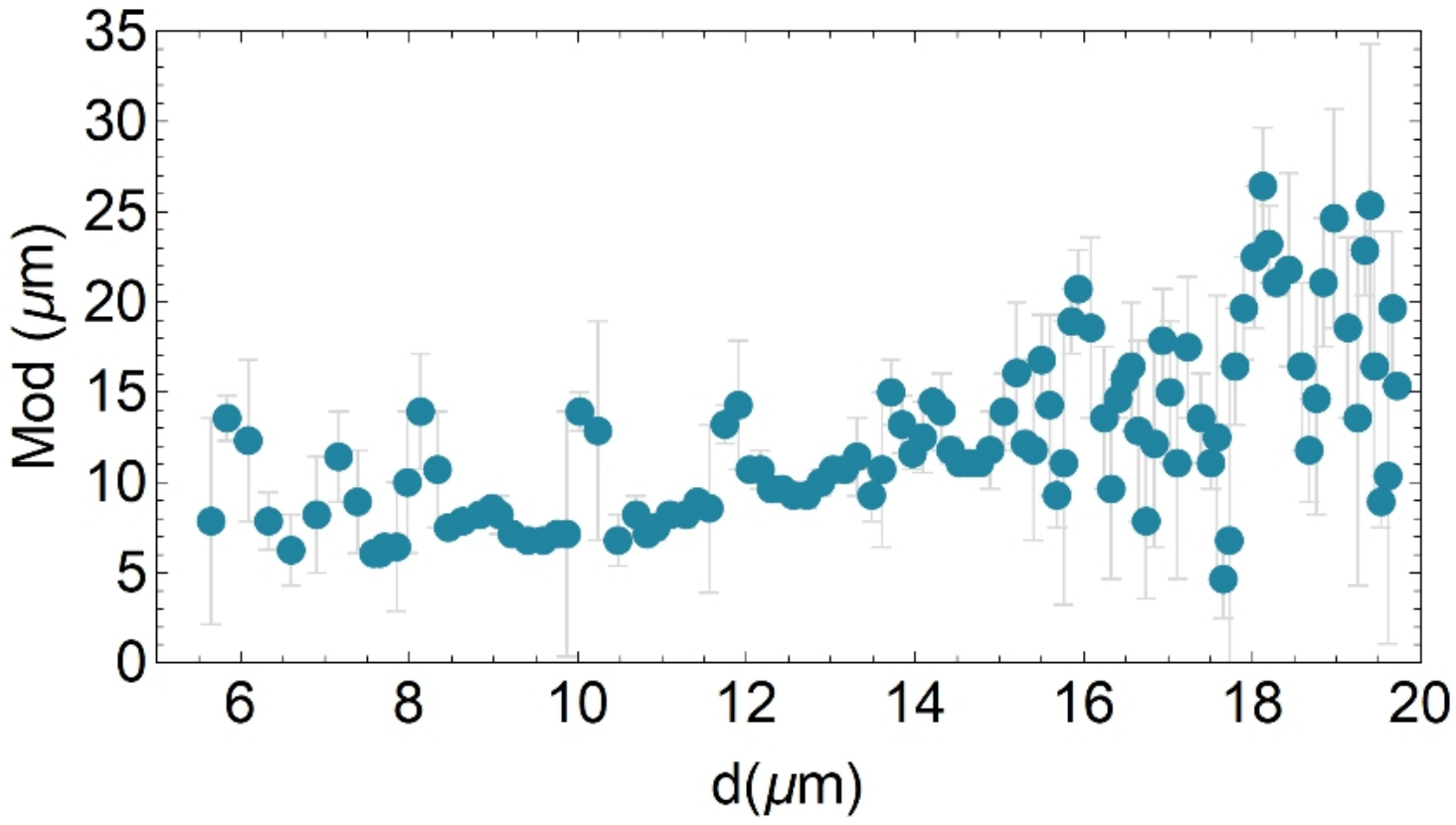

Supporting Figure 7. Macroscopic structure periodicity as a function of thickness in the $N_{TBF}$* phase as determined in the wedge cell (EHC).

## Supporting material Modulated Instabilities

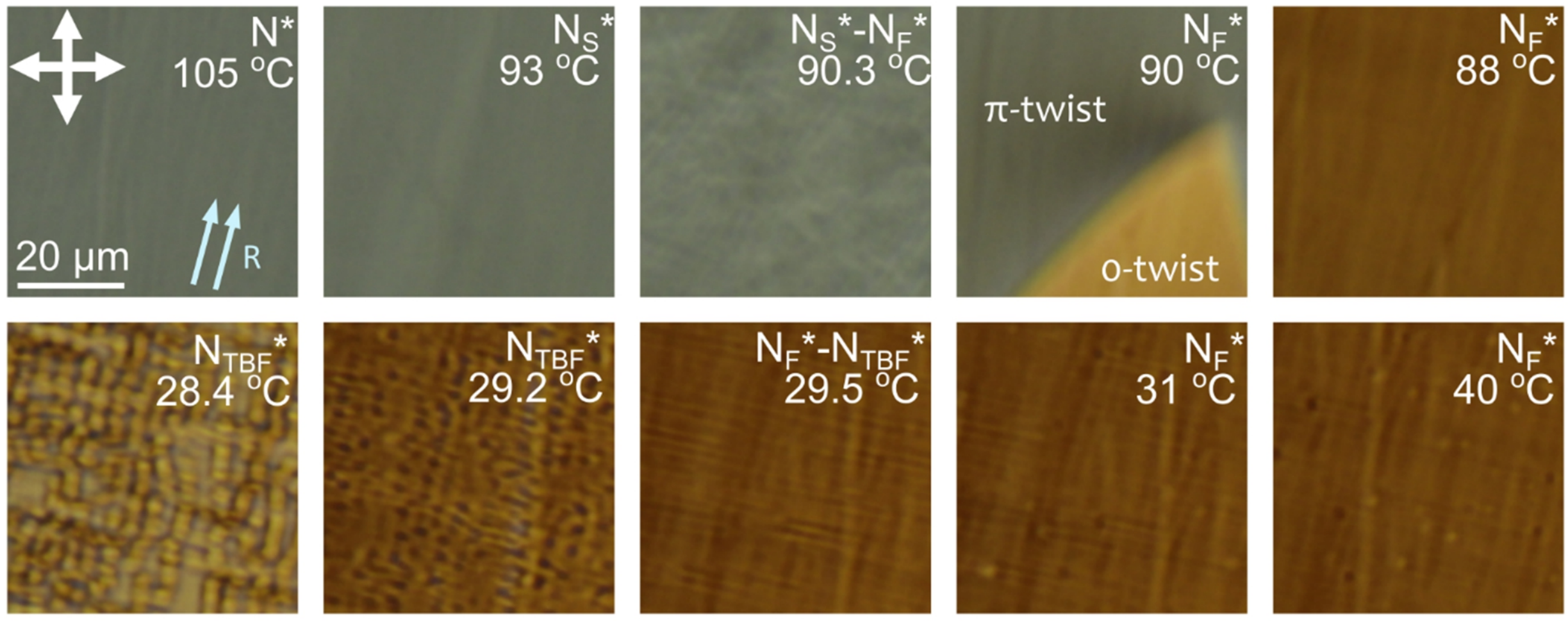

Supporting Figure 8. Polarizing optical microscopy images of F7* sample confined in a 2 μm parallel rubbed cell around the $N_F$-$N_{TBF}$ transition.

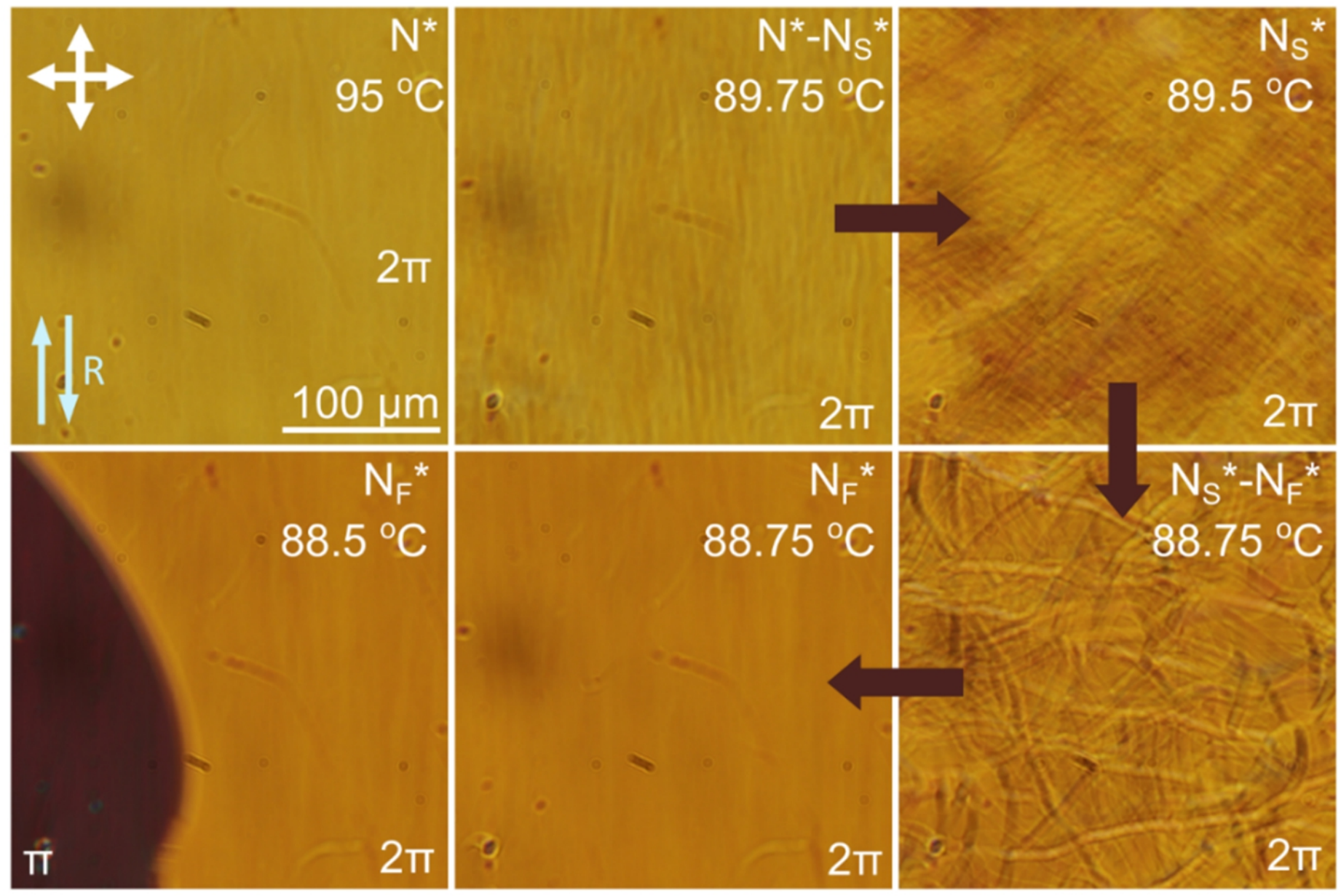

Supporting Figure 9. Formation of a π-domain for F7+S811-1 in a 5 µm antiparallel rubbed cell on cooling from the N* phase as detected with POM. The thickness of 5 µm, the pitch, and the aligning conditions stabilize a hypo-twisted π-domain in the $N_F$* phase.

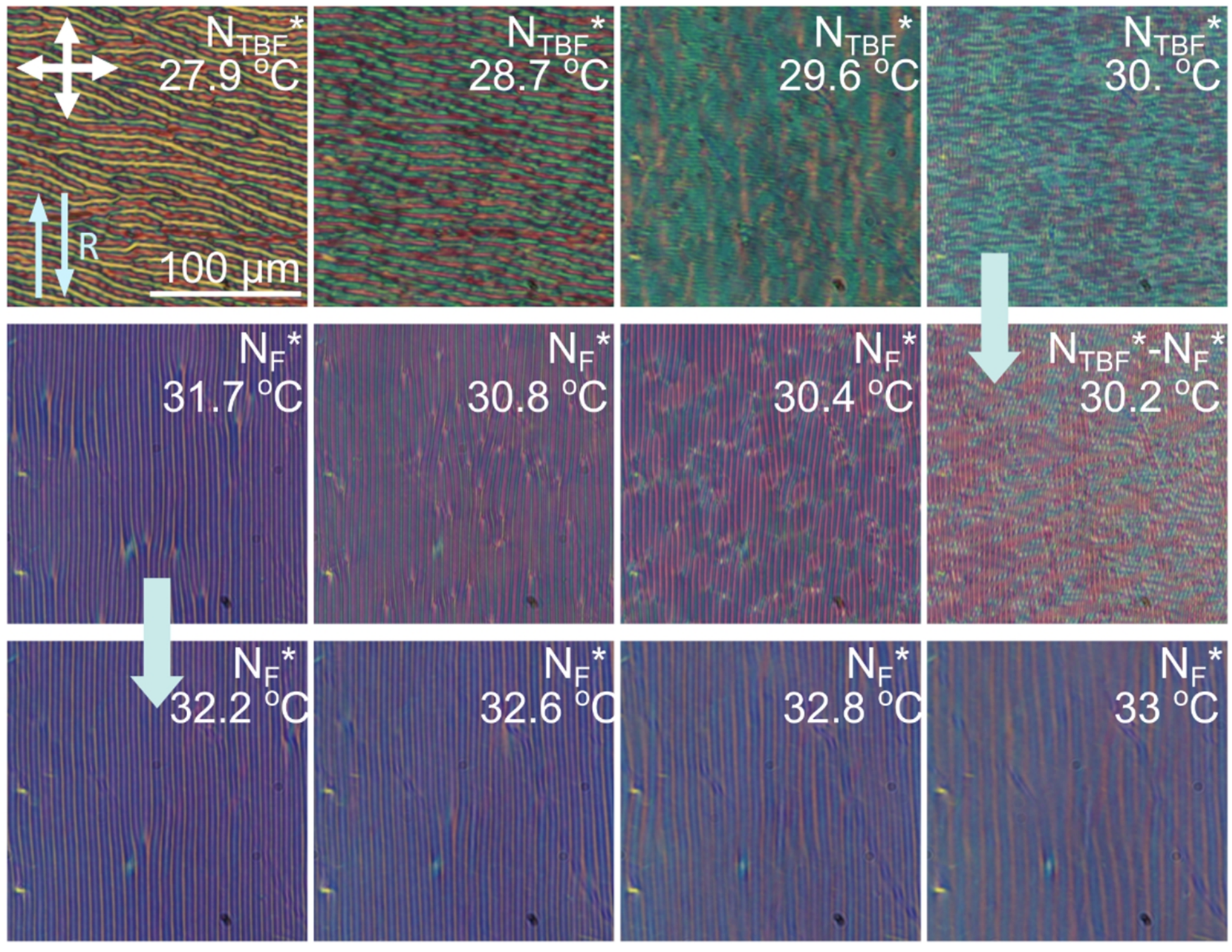

Supporting Figure 10. Polarizing optical microscopy images of the $N_F$*-$N_{TBF}$* transition on heating for F7* (F7+S811-1%) in a 5 µm antiparallel rubbed cell.

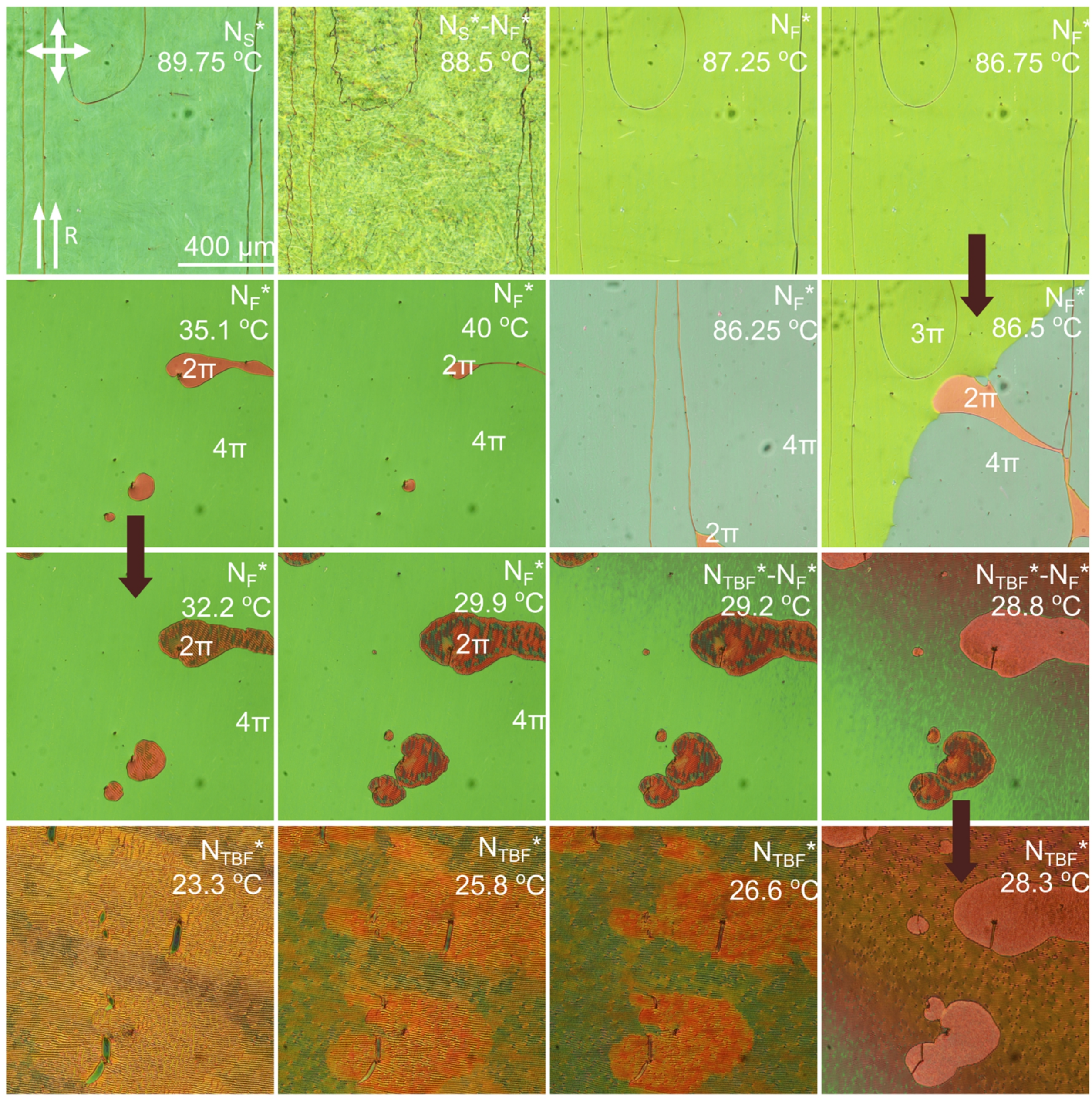

Supporting Figure 11. Formation of 2π and 4π -domains for F7* (F7+S811-1%) in a 10 μm parallel rubbed cell on cooling from the N* phase as seen by POM. The thickness, pitch and aligning conditions stabilize a hypertwisted 4π-domain in the $N_F$* phase. Small islands of 2π-twisted structures can also be observed. Withing these hypotwisted regions instabilities occur right before the transition to the low temperature $N_{TBF}$* phase. Stabilizing the sample at this temperature, as well as cycling through the $N_F$*-$N_{TBF}$* transition promotes the growth of the 2π-domains. (See Supporting Figure 11) This agrees with the observed tendency for pitch increase before the transition, which would favor 2π over 4π structures.

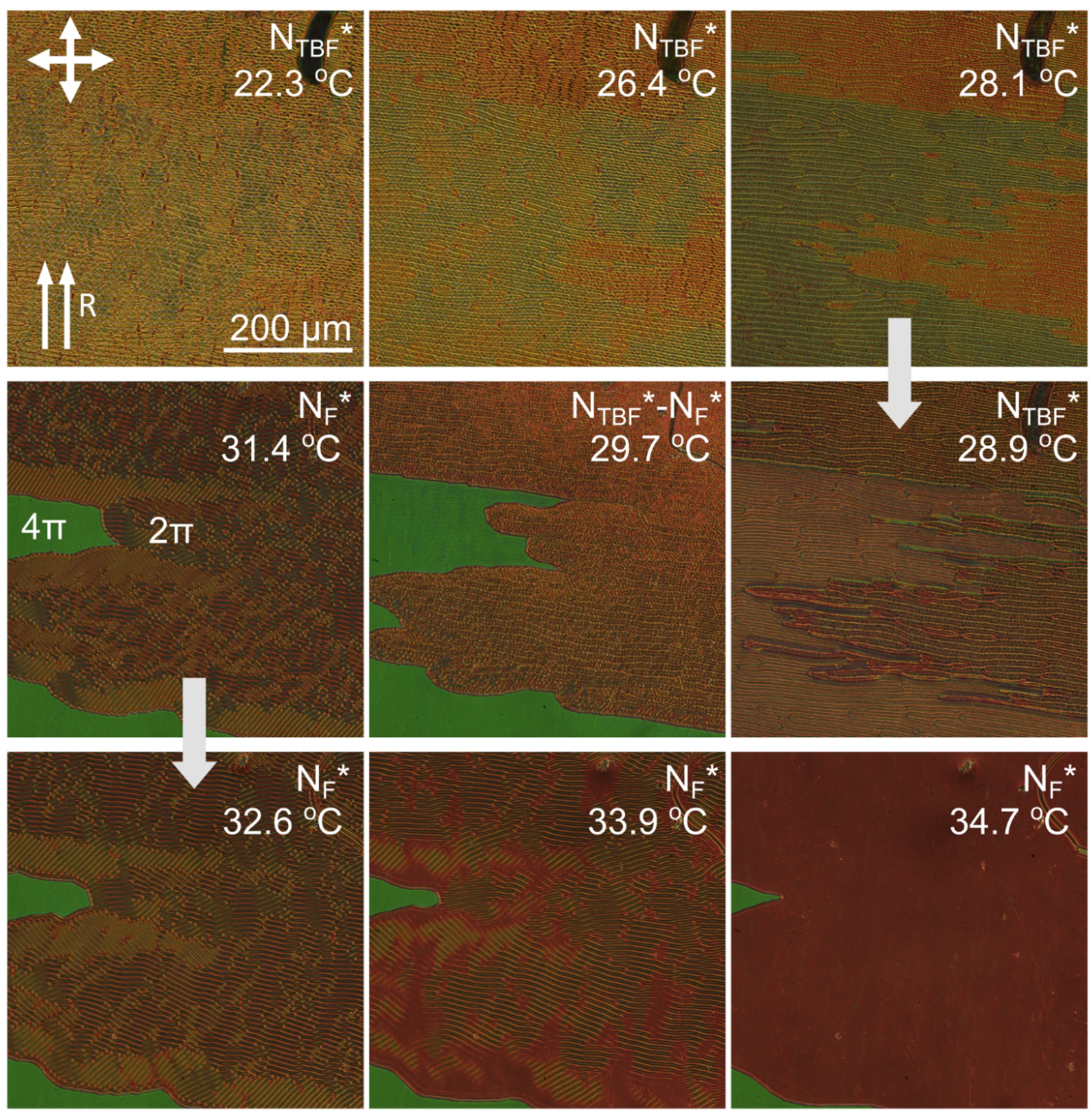

Supporting Figure 12. Polarizing optical microscopy images of the $N_F$*-$N_{TBF}$* transition on heating for F7* (F7+S811-1%) in a 10 µm parallel rubbed cell showing, as mentioned before, the stabilization of 2π-domains versus 4π-domains.

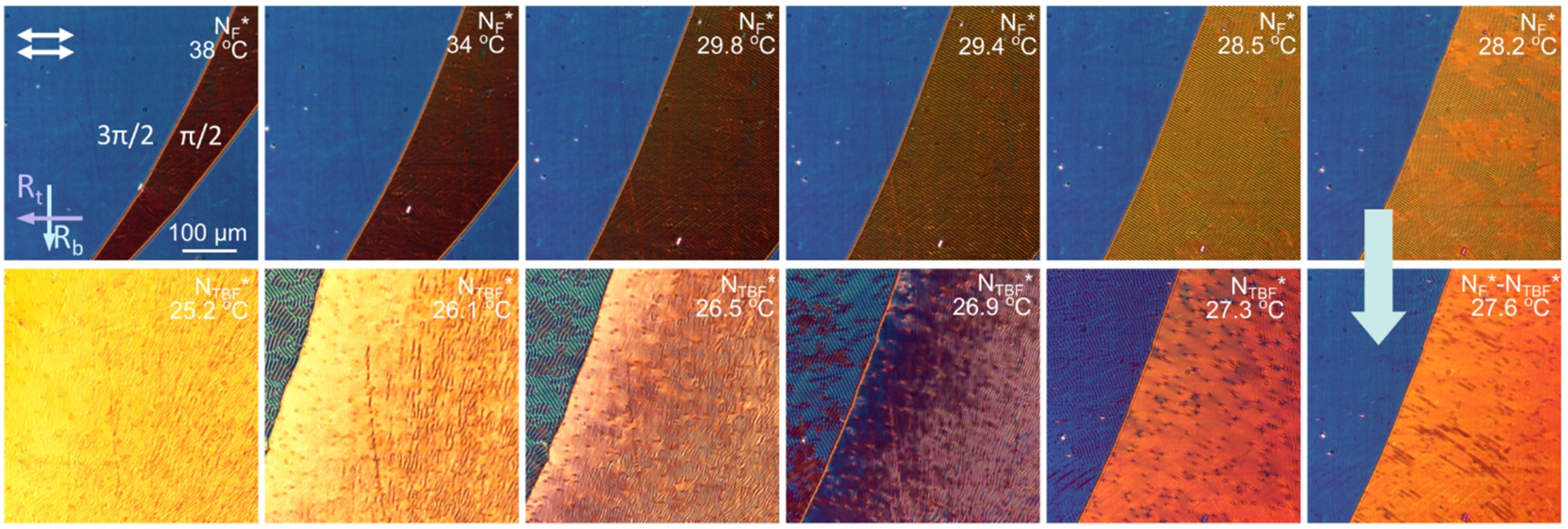

Supporting Figure 13. Polarizing optical microscopy images of the $N_F$*-$N_{TBF}$* transition on heating for F7* (F7+S811-1%) in a 5 μm left-handed 90-twisted rubbed cell, showing the different behaviour of the elastic instabilities in the π/2 and 3π/2 twisted areas. On cooling from the N* phase, 3π/2-twist structure is stabilized at high temperatures in the $N_F$* phase. However, at lower temperatures π/2-twsited areas start to grow due to the slight increase of the pitch.

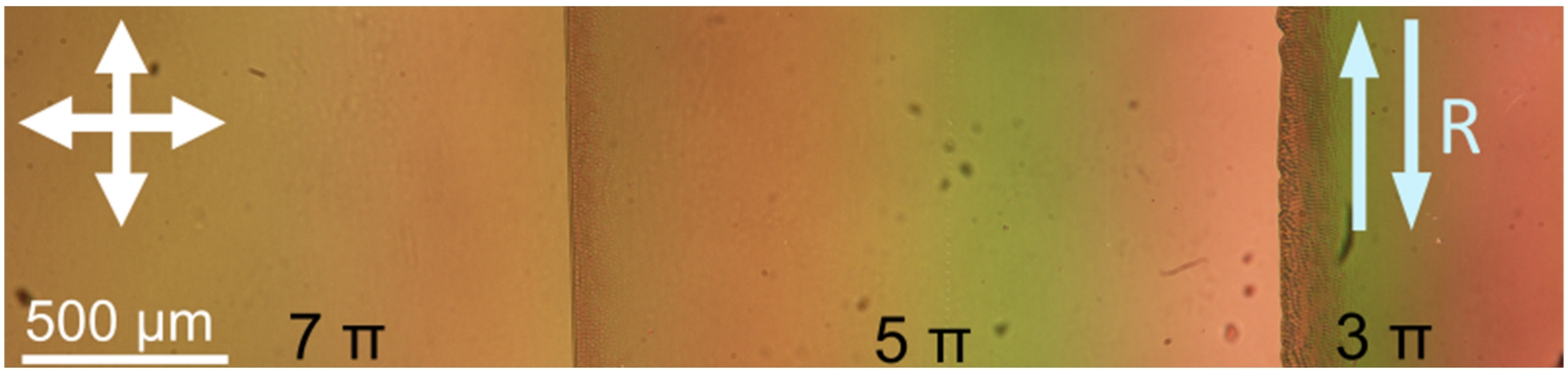

Supporting Figure 14. Overview of custom-made wedge cell with antiparallel rubbing showcasing the 3π, 5π, and 7π regions as imaged in POM in the $N_F$* phase at 29.7 °C. Double headed arrows indicate the position of crossed polarizers, and light-blue arrows indicate the rubbing direction in the confining cell surfaces.

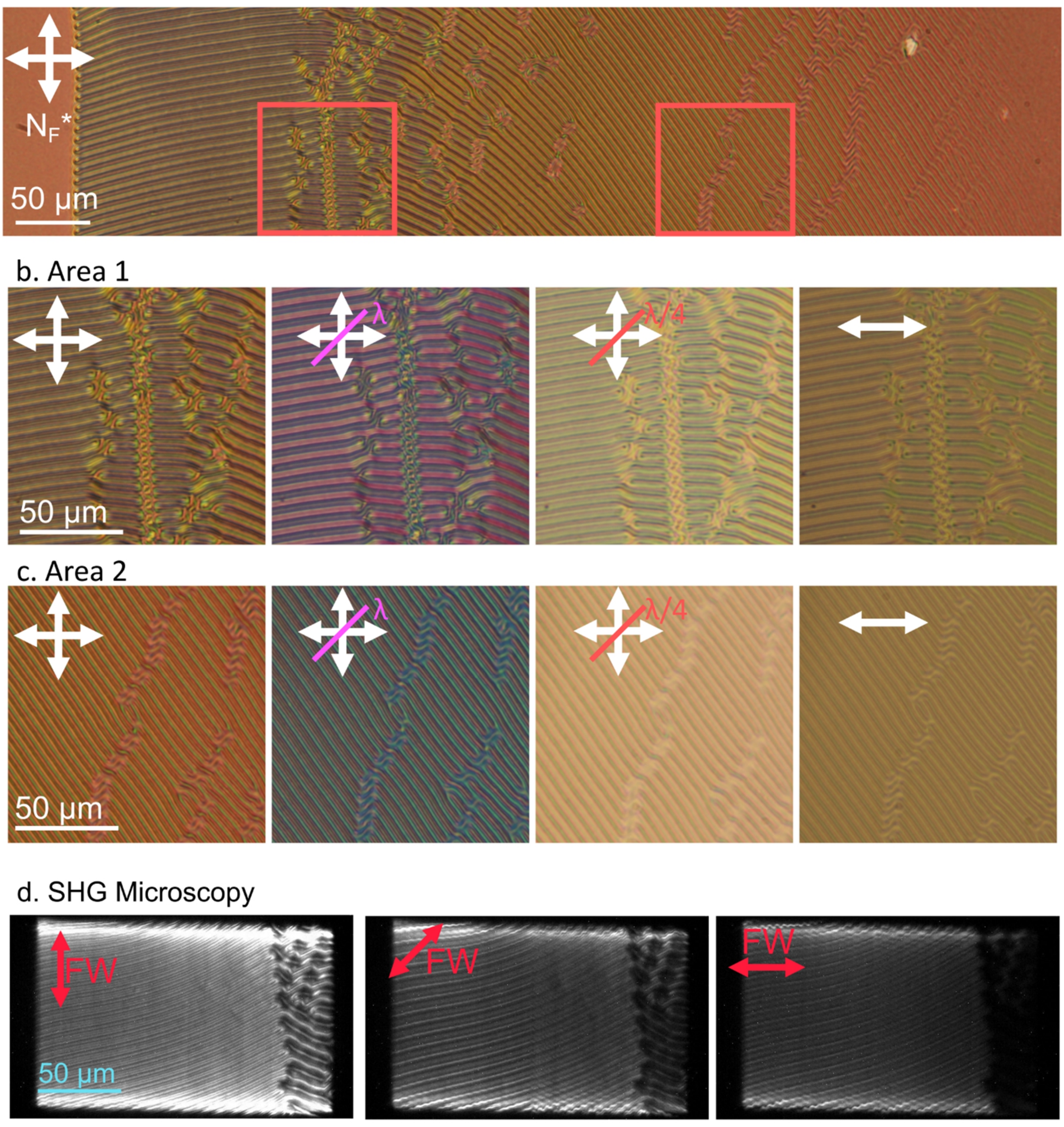

Supporting Figure 15. (a) Polarizing optical microscopy images of the 5π-3π region boundary in the $N_F$* phase at 29.5 °C. The top image shows an overview across different cell thicknesses. (b-c) Zoom-in images of areas marked in (a) by pink-squares at different conditions: crossed polarisers, crossed polarisers and full-lambda plate, crossed polarisers and quarter-lambda plate, and only polariser. (d) SHG-Microscopy images of the thicker 3π region section at different incoming fundamental wave polarisations and no analyser.

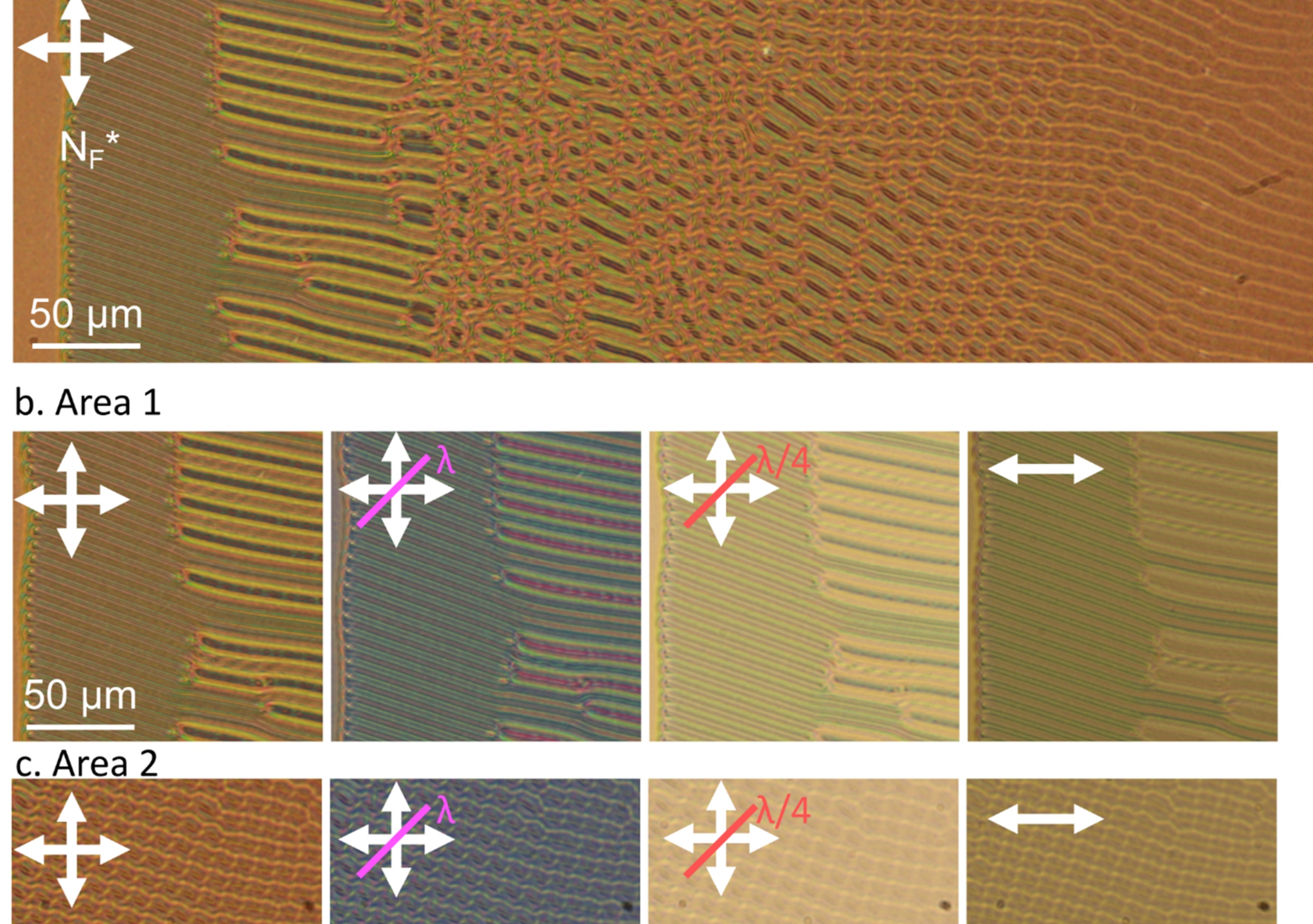

d. SHG-Microscopy

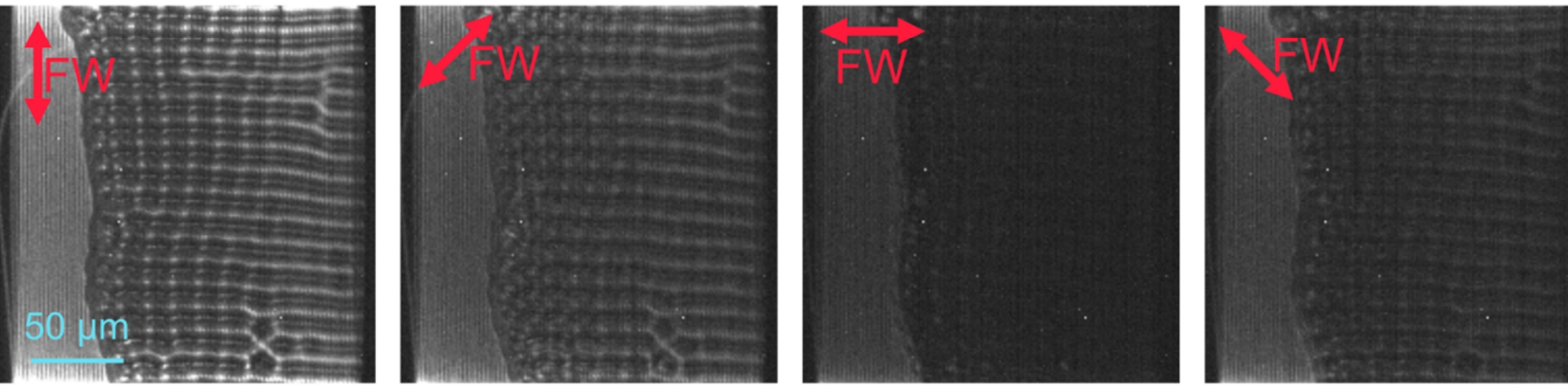

Supporting Figure 16. (a) Polarizing optical microscopy images of the 7π-5π region boundary in the $N_F$* phase at 29.5 °C. Top image shows an overview across different cell thicknesses. (b-c) Zoom-in images of two different areas from (a) at different conditions: crossed polarisers, crossed polarisers and full-lambda plate, crossed polarisers and quarter-lambda plate and only polariser. (d) SHG-Microscopy images of the thicker 5π region section at different incoming fundamental wave polarisations and no analyser.

## Supporting material modelling of instabilities

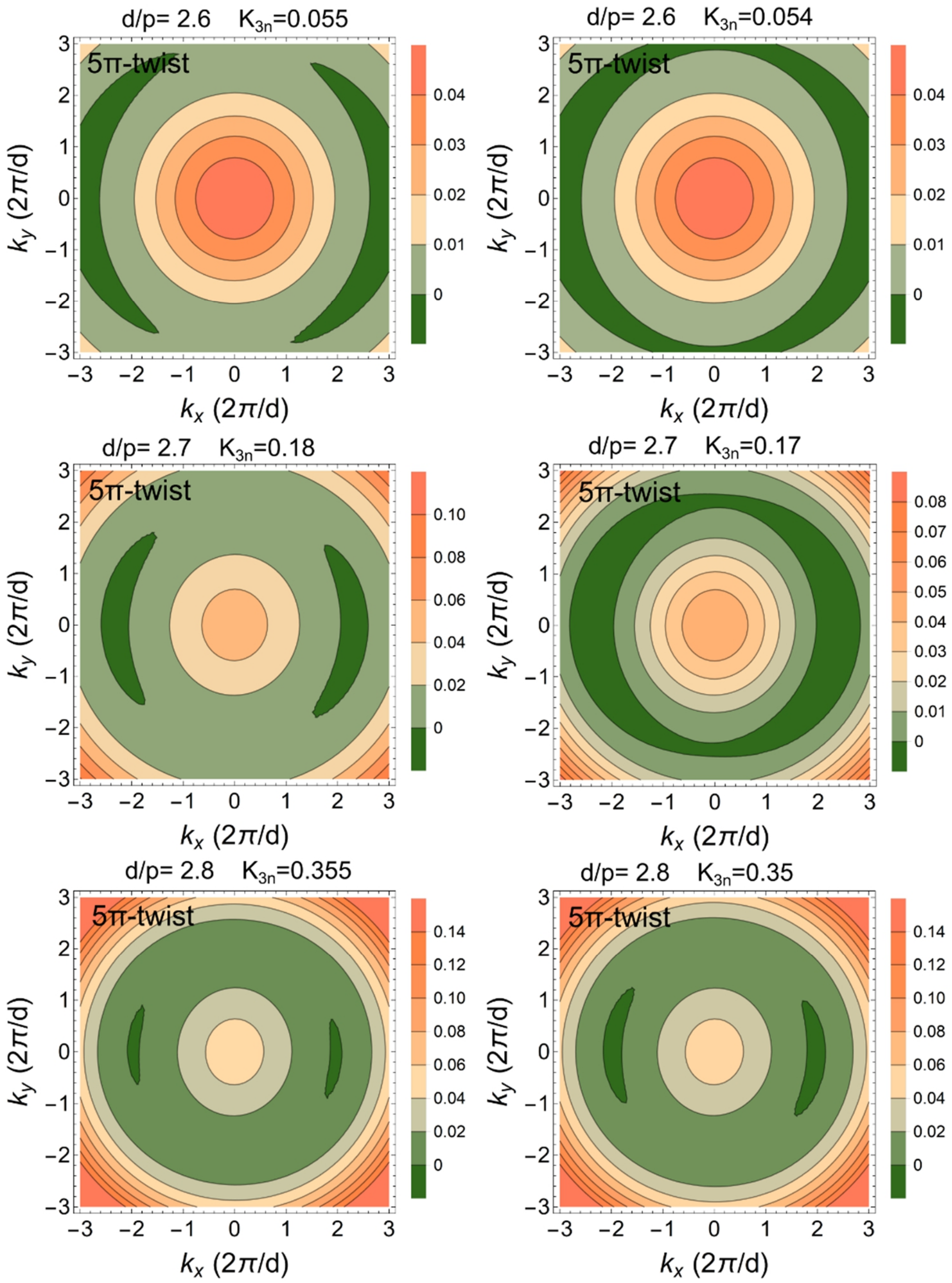

Supporting Figure 17. Two-dimensional maps of the lowest calculated eigenvalues as a function of the instability wavevector ($k_x$,$k_y$) for different confinement ratios and $K_{3n}$=$K_3$/$K_2$ for the 5π-twisted structures. Results show that for slightly lower $K_{3n}$ ratios than at the onset, the dispersion promptly exhibits a shallow, round minimum, allowing multiple lattice symmetries, as experimentally observed.

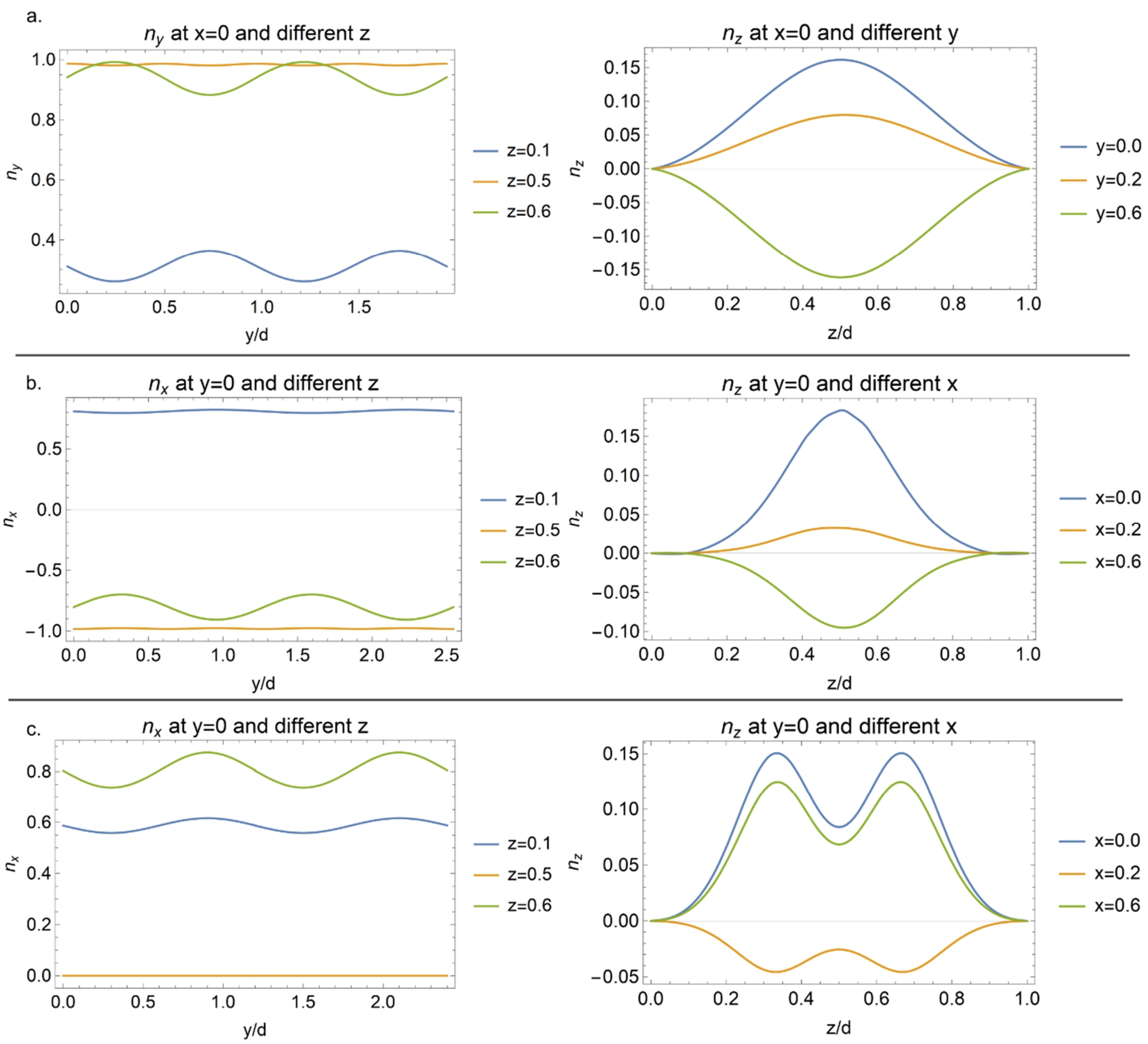

Supporting Figure 18. Profiles of the $n_x$, $n_y$ and $n_z$ components of the director field for the structures plotted in the manuscript Figure 6 for a) 1π-, b) 2π- and c) 3π-twisted structures. Left plots show the $n_x$($n_y$) component as a function of x(y) at different positions along the cell thickness (z). The right plots show the $n_z$ component of the director field as a function of z at different x(y) coordinates.

## Description of Supporting Movies

Movie S1. Transition from the $N_F$* phase to the $N_{TBF}$* phase as observed by polarizing optical microscopy (POM) in a 5 µm parallel rubbed cell. The sample is placed between crossed polarizers, in vertical and horizontal directions, with the cell rubbing direction rotated around 20 degrees. The experiment was performed on cooling at 0.3 °C/min, which in the movie corresponds to 0.4 °C/s. Movie starts at 33 °C. Full movie width corresponds to 725 µm.

Movie S2. Temperature evolution of Grandjean-Cano texture in a wedge cell ($\theta = 0.68°$) with parallelly rubbed surfaces. The temperature starts at 110 °C. The experiment was performed on cooling at 0.5 °C/min, which in the movie corresponds to 2 °C/s. The sample is placed between crossed polarizers, in vertical and horizontal directions. The LC cell has parallel rubbing directions on the surface, which in the movie correspond to the horizontal direction, i.e., along the cell wedge. The thickness of the cell varies from the thinner edge (right) to the thicker (left). Full movie width corresponds to 3.5 mm.

Movie S3. Temperature evolution of the instabilities in the π-twisted structure (5.2 µm antiparallel rubbed cell) prior to the transition from the $N_F$* phase to the $N_{TBF}$* phase as observed by POM (left) and the corresponding 2D Fourier transform (Right). The experiment was performed on cooling at 0.5 °C/min. The sample is placed between crossed polarizers, in vertical and horizontal directions. The LC cell has antiparallel rubbing directions on the surface, which in the movie correspond to the vertical direction (as shown in Figure 3 in the manuscript). Full movie width corresponds to 125 µm.

Movie S4. Temperature of the instabilities in the π-twisted structure (5.2 µm antiparallel rubbed cell) prior to the transition from the $N_F$* phase to the $N_{TBF}$* phase as observed by SHG-Microscopy. The experiment was performed on cooling at 0.5 °C/min, acquiring one image every 24 seconds. The video corresponds to the temperature range from 31.5°C to 24°C, i.e. temperature rate 0.2 °C/s. Incoming 800 nm fundamental wave is polarized along cell rubbing direction, i.e. vertical in the image. No analyser is employed for image recording.